\documentclass[letter,11pt]{article}

\newcommand{\ket}[1]{\ensuremath{\left| #1 \right\rangle}}

\newcommand{\br}[1]{\ensuremath{\left\langle #1 \right.}}

\newcommand{\bra}[1]{\ensuremath{\left. \br{#1} \right|}}

\newcommand{\bk}[2]{\br{{#1}}\ket{{#2}}}

\newcommand{\kb}[2]{\ket{{#1}}\bra{{#2}}}

\newcommand{\ip}[3]{\bra{{#1}} {#2} \ket{{#3}}}

\newcommand{\ev}[2]{\bra{{#1}} {#2} \ket{{#1}}}

\newcommand{\proj}[1]{\kb{{#1}}{{#1}}}

\newcommand{\trace}[1]{\ensuremath{\mathrm{Tr}\left[{#1}\right]}}

\newcommand{\partrace}[2]{\ensuremath{\mathrm{Tr}_{{#2}}\left[{#1}\right]}}

\newcommand{\mean}[1]{\ensuremath{\left\langle {#1} \right\rangle}}
\newcommand{\magn}[1]{\ensuremath{\left| {#1} \right|^2}}
\newcommand{\func}[1]{\ensuremath{\left[ {#1} \right]}}
\newcommand{\mod}[1]{\ensuremath{\left| {#1} \right|}}

\newcommand{\com}[2]{\ensuremath{\left[{#1},{#2}\right]_{-}}}

\newcommand{\partime}[1]{\ensuremath{\frac{\partial {#1}}{\partial t}}}
\newcommand{\partxy}[2]{\ensuremath{\frac{\partial {#1}}{\partial {#2}}}}

\newcommand{\eqnwrap}{\nonumber \\ &&}

\newcommand{\basis}[1]{\ensuremath{\left\{\ket{#1}\right\}}}

\begin{document}
\title{The Physical Basis of the Gibbs-von Neumann entropy.}
\author{O J E Maroney\\
Perimeter Institute for Theoretical Physics \\
31 Caroline Street North,
Waterloo,
Ontario, N2L 2Y5, Canada\\
omaroney@perimeterinstitute.ca\\
pi-foundqt-36}
\date{January 21, 2008}
\maketitle

\begin{abstract}
We develop the argument that the Gibbs-von Neumann entropy is the
appropriate statistical mechanical generalisation of the
thermodynamic entropy, for macroscopic and microscopic systems,
whether in thermal equilibrium or not, as a consequence of
Hamiltonian dynamics.  The mathematical treatment utilises well
known results \cite{Gib1902,Tol1938,Weh78,Par89b}, but most
importantly, incorporates a variety of arguments on the
phenomenological properties of thermal states
\cite{Szi1925,TQ1963,HK1965,GB1991} and of statistical
distributions\cite{GH1976b,PW1978,Lenard1978}. This enables the
identification of the canonical distribution as the unique
representation of thermal states without approximation or
presupposing the existence of an entropy function. The Gibbs-von
Neumann entropy is then derived, from arguments based solely on the
addition of probabilities to Hamiltonian dynamics.
\end{abstract}

\tableofcontents

\newpage

\section{Introduction}\label{s:introduction}
\begin{quote}The laws of statistical mechanics apply to conservative
systems of any number of degrees of freedom, and are
exact.\cite{Gib1902}\end{quote}

The statistical mechanics considered by Gibbs, in his classic
treatise of 1902, is a more general structure than thermodynamics.
It applies to any kind of Hamiltonian system in which probabilistic
reasoning is valid.  Useful properties may be derived without any
reference to thermodynamic quantities, and can be used without any
consideration of whether or not one is dealing with a thermal
system.

Nevertheless, thermal systems exist, are important and it is
necessary that statistical mechanics gives an account of them,
including the phenomena usually described by thermodynamics.  Making
this connection is surprisingly hard, without introducing ``question
begging'' assumptions.  Gibbs tentatively attempted to make this
connection, in \cite{Gib1902}[Chapter XIV], but only referred to
thermodynamic ``analogies''.

Criticisms of the Gibbs approach, and the Gibbs entropy as a
\textit{ thermodynamic} entropy are not hard to find these
days\cite{Cal99,She99a,Gol2001,Alb01} and go back at least as far
as\cite{EE1912}.  The principal purpose of this paper, is to argue
that the Gibbs-von Neumann entropy\footnote{We will be working with
quantum mechanical systems, so will derive the von Neumann entropy.}
can be derived, from physical arguments, without problematical
assumptions, as precisely what one should desire for a \textit{
statistical} generalisation of \textit{ thermodynamic} entropy.

The method will not be to directly attempt to find statistical
mechanical term to act as a thermodynamic entropy.  There is no
uncontroversial definition of entropy, outside of classical
phenomenological equilibrium thermodynamics, but the world is not in
equilibrium and statistical fluctuations occur.  Instead the
approach will be to develop statistical mechanics as a broader
subject than thermal physics. When trying to apply statistical
mechanics to thermal phenomena we will consider some basic physical
properties of thermal states and then apply those properties to
statistical mechanics.

Of central importance, and what differs most from more traditional
treatments, will be the justification for the derivation of the
canonically distributed density matrix as the unique statistical
distribution that can represent thermal states.  Here the key
arguments will be Szilard's 1925 derivation of the canonical
distribution\cite{Szi1925} from phenomenological grounds, the
concept of passive distributions\cite{PW1978,Lenard1978,Sewell1980}
and their relationship to the adiabatic availability of
energy\cite{HK1965,GH1976b,GB1991}. It is an interesting feature of
these, that it does not depend upon whether the system is large or
small or in thermal equilibrium or not.  It applies to any situation
in which the use of probability distributions are valid.

We will first explore the mathematical structure of Gibbs
statistical mechanics, as applied to quantum mechanics.  The
mathematics will, in a large part be
recognisable\cite{Gib1902,Tol1938,Weh78,Par89b}, but the emphasis
will be to show what can (and cannot) be derived \textit{ without}
making physical assumptions.  The section will, of necessity, appear
rather abstract and unmotivated.

The structure so developed produces equations very like those that
occur in thermodynamics. Actually connecting these equations with
thermodynamics requires a logical jump which appears to assume
precisely the thing which one seeks to justify.  We will briefly
review why this is so and some of the attempts to make this jump.

We then return to the physical basis of the statistical approach.
The statistical approach applies whenever it is meaningful to use
probabilities.  Deciding when this is so is not uncontroversial, but
we will not address that problem here.  Instead we will explore what
the consequences are when a probabilistic description is meaningful.
A remarkably large amount of the familiar structure of statistical
mechanics can be derived without any reference to thermal concepts
or notions of entropy.  Particular attention will be drawn to the
closely related concepts of adiabatic availability and passive
distributions.

Only after we have derived the general structure of quantum
statistical mechanics will we consider thermal systems.  To examine
what are the statistical mechanics of thermal systems, we first need
to identify what we mean by a thermal system.  We identify four
physical properties, which we suggest are observed properties of
thermal states.  These properties uniquely select the canonical
probability distribution.  If this argument is accepted, we then
proceed in well established steps to develop thermal heat baths, the
temperature scale and finally the form of a statistical mechanical
entropy from physically motivated arguments.

These physical arguments are valid for systems of any size, for
non-equilibrium systems as well as for equilibrium systems, indeed
for any situation where the use of a probability distribution is
physically justified.

\section{Mathematical formalism}\label{s:distributions}
We establish the properties of a particular type of function, which
for want of a better word we shall call a distribution, on the state
space of a system that has a Hamiltonian evolution.  No physical
interpretation is placed upon either this type of function nor of
the derived properties.  The object is to establish exactly which
purely mathematical properties can be defined without needing to
introduce physical justifications.

This section will, perhaps, seem needlessly abstract and physically
unmotivated.  This is quite correct!  We develop the mathematics
first to ensure that no physical assumptions have been used in their
derivation.  This is try to avoid circular reasoning when we come to
consider the appropriate descriptions of physical processes.

When we do start to identify physical processes with mathematical
structures, we wish to be clear which properties legitimate that
identification, which properties then follow directly from that
identification, and which properties require further assumptions or
justification. Readers who are prepared to take this on trust may
jump directly to Section \ref{s:probabilities} where we will start
considering the properties of physical systems.

\subsection{Distributions}\label{ss:distributions}
The quantum mechanical state space is a Hilbert space $\Pi$ and has
a Hamiltonian evolution operator, $H(t)$. For the purposes of this
paper, a distribution on the state space is an operator $\Omega$ on
the state space, with orthonormal eigenstates $\basis{\beta}$ and
real eigenvalues $\omega_\beta$, for which:
\begin{eqnarray}
\Omega&=&\sum_\beta \omega_\beta \proj{\beta}\\
\omega_\beta & \geq & 0 \\
\sum_\beta \omega_\beta &=& 1 \\
i \hbar\partime{\Omega}&=&\com{H}{\Omega}
\end{eqnarray}
Both the eigenstates and eigenvalues may be evolving in time due to
$H$.  We may also write the time evolution of the distribution in
the unitary form:
\begin{equation}
\Omega(t-t_0)=U\Omega(t_0)U^\dag
\end{equation}
where $U$ is the solution of the operator equation
\begin{equation}
i \hbar \partime{U}=HU
\end{equation} or in the more general superoperator form:
\begin{equation}
\Omega(t-t_0)=L(t-t_0)\func{\Omega(t_0)}
\end{equation}
We will refer to the combination of a state space $\Pi$, Hamiltonian
evolution $H$ on the state space, and distribution $\Omega$ over the
state space as a system.
\subsubsection{Subdistributions}\label{ss:subdistributions} A
subdistribution is the normalised portion of a distribution that is
non-zero over a restricted region $R \subset \Pi$ of the state
space:
\begin{equation}
\Omega_i=\sum_{\beta \in R} \frac{\omega_\beta}{\sum_{\beta \in R}
\omega_\beta} \proj{\beta}
\end{equation}
Two subdistributions are non-overlapping if there is no region of
the state space for which they are both non-zero:
\begin{equation}
\Omega_i \Omega_j = \delta_{ij} (\Omega_i)^2
\end{equation}
A distribution may be decomposed into non-overlapping
subdistributions:
\begin{equation}
\Omega=\sum_i w_i \Omega_i
\end{equation}
It will be useful to do this by constructing a complete set of
non-overlapping projectors, $K_i$, such that
\begin{eqnarray}
K_i K_j &=& \delta_{ij} K_i \\
\sum_i K_i &=& I
\end{eqnarray}
constructed from the eigenstates of the distribution:
\begin{eqnarray}
\Omega&=&\sum_{\alpha} \omega_{\alpha} \proj{\alpha} \\
\bk{\alpha}{\alpha^{\prime}}&=&\delta_{\alpha \alpha^{\prime}} \\
K_i&=&\sum_{\alpha \subset i} \proj{\alpha}\\
\Omega_i&=&\frac{K_i \Omega K_i}{\trace{K_i \Omega K_i}} \\
w_i&=&\trace{K_i \Omega K_i}
\end{eqnarray}
\subsubsection{Subspaces}\label{ss:subspaces} When the Hilbert space
can be separated into a product of two subspaces $\Pi=\Pi_1\otimes
\Pi_2$ we form
 the marginal distributions
\begin{eqnarray}
\overline{\Omega_1}&=&\partrace{\Omega}{2} \\
\overline{\Omega_2}&=&\partrace{\Omega}{1}
\end{eqnarray}
The marginal distributions do not generally evolve by a Hamiltonian
evolution, but the evolution may still be expressed by a
superoperator equation:
\begin{equation}
\overline{\Omega_1(t)}=L_1\func{\overline{\Omega_1(0)}}
\end{equation}
\subsection{Operators}\label{ss:operators}
Given an operator $A(t)$, on the state space, we may define the
value of that operator for the distribution $\Omega(t)$ by:
\begin{equation}
\mean{A(t)}_{\Omega(t)} = \trace{A(t)\Omega(t)}
\end{equation}
Given the Hamiltonian evolution operator $H(t)$, for the state
space, then
\begin{equation}
i \hbar
\partime{\mean{A(t)}_{\Omega(t)}}=i
\hbar\mean{\partime{A(t)}}_{\Omega(t)}+\mean{\com{A(t)}{H(t)}}_{\Omega(t)}
\end{equation}
If the system is not isolated, then it is a subsystem $\Pi_1$ of a
larger space $\Pi=\Pi_1 \otimes \Pi_2$.  The Hamiltonian may be
rewritten as the sum of three terms:
\begin{enumerate} \item A term operating solely upon subsystem 1, $H_1(t)$;
\item A term operating solely upon subsystem 2, $H_2(t)$;
\item and a term operating jointly as an interaction between the two systems
$V_{12}(t)$.
\end{enumerate}
\begin{equation}
H(t)=H_1(t)\otimes I_2 +I_1 \otimes H_2(t) +V_{12}(t)
\end{equation}
Now the evolution of the marginal distribution
$\overline{\Omega_1(t)}=\partrace{\Omega(t)}{2}$ will not, in
general, be describable by a Hamilton evolution operator.

If we take an operator $A_1(t)$ that acts solely upon the space of
the subsystem $\Pi_1$, we find:
\begin{equation}
i \hbar \partime{\mean{A_1(t)\otimes I_2}_{\Omega(t)}}= i
\hbar\mean{\partime{A_1(t)}}_{\overline{\Omega_1(t)}}+
\mean{\com{A_1(t)}{H_1(t)}}_{\overline{\Omega_1(t)}}
+\mean{\com{A_1(t)\otimes I_2}{V_{12}(t)}}_{\Omega(t)}
\end{equation}

Unless $\com{A_1(t)\otimes I_2}{V_{12}(t)}=0$ we appear to have a
dependancy upon the full distribution $\Omega(t)$.  To eliminate
this we express the evolution of the marginal distribution through
the evolution of it's eigenstates and eigenvalues:
\begin{equation}
\overline{\Omega_1(t)}=\sum_\alpha \omega_\alpha(t)\proj{\alpha(t)}
\end{equation}
The set of eigenstates $\basis{\alpha(t)}$ will always be a basis
for the subspace, so there exists an unitary operator $\Upsilon(t)$
for which
\[
\ket{\alpha(t)}=\Upsilon(t)\ket{\alpha(0)}
\]
and whose evolution is generated by a Hamiltonian operator
$\Theta(t)$:
\[
i \hbar \partime{\Upsilon(t)}=\Theta(t) \Upsilon(t)
\]
This gives
\begin{equation}
i \hbar
\partime{\mean{A_1(t)}_{\overline{\Omega_1(t)}}}= i
\hbar\mean{\partime{A_1(t)}}_{\overline{\Omega_1(t)}}+ \sum_\alpha i
\hbar \ev{\alpha(t)}{A_1(t)}\partime{\omega_\alpha(t)}+
\mean{\com{A_1(t)}{\Theta(t)}}_{\overline{\Omega_1(t)}}
\label{eq:workheat}
\end{equation}
which is expressed purely in terms of operators upon, and a
distribution over, the subspace.

Note that if any of the commutators $\com{A_1(t)}{\Theta(t)}$,
$\com{\overline{\Omega_1(t)}}{A_1(t)}$ or
$\com{\Theta(t)}{\overline{\Omega_1(t)}}$ are zero, the third term
disappears to give:
\begin{equation}
\partime{\mean{A_1(t)}_{\overline{\Omega_1(t)}}}=
\mean{\partime{A_1(t)}}_{\overline{\Omega_1(t)}}+ \sum_\alpha
\ev{\alpha(t)}{A_1(t)}
\partime{\omega_\alpha(t)}
\end{equation}

The following two terms will be useful later on:
\begin{eqnarray}
\Delta A_1(t)&=&\int_0^{t}
\partime{\mean{A_1(t)}_{\overline{\Omega_1(t)}}} dt =\mean{A_1(t)}_{\overline{\Omega_1(t)}}-\mean{A_1(0)}_{\overline{\Omega_1(0)}} \\
D\func{A_1(t)}&=&\int_0^{t}
\mean{\partime{A_1(t)}}_{\overline{\Omega_1(t)}} dt
\end{eqnarray}

\subsection{Gibbs-von Neumann measure}\label{ss:gibbs}
We now introduce the Gibbs-von Neumann measure of a distribution:
\begin{equation}
G\func{\Omega}=\trace{\Omega \ln \func{\Omega}}
\end{equation}
There may be the perception that we have introduced a ``question
begging'' step, as to why we introduce this particular measure. We
suggest that this is not the case, as we have made no physical
interpretation of this measure.  We introduce it simply to establish
some of its mathematical properties, devoid of any interpretation.
\subsubsection{Concavity}\label{sss:convex}
The Gibbs-von Neumann measure is a concave function and this has the
property, that given any two distributions $\Omega$ and
$\Omega^\prime$, then
\begin{equation}
\trace{\Omega \left(\ln \func{\Omega}-\ln
\func{\Omega^\prime}\right)} \geq 0
\end{equation}

\subsubsection{Subspaces}\label{sss:gibbssubens}
When a space can be separated into two subspaces $\Pi=\Pi_1 \otimes
\Pi_2$, we can define a measure of the correlation of the
distribution between the subspaces as:
\begin{equation}
C\func{\Omega}=G\func{\overline{\Omega_1}}+G\func{\overline{\Omega_2}}-G\func{\Omega}
\geq 0 \label{eq:correlation}
\end{equation}
Equality occurs if, and only if, the systems are uncorrelated
$\Omega=\overline{\Omega_1} \otimes \overline{\Omega_2}$.

This has a direct consequence for the evolutions of initially
uncorrelated systems that are allowed to interact.  If the systems
are uncorrelated at $t=0$, so that $\Omega(0)=\Omega_1(0)\otimes
\Omega_2(0)$ but allowed to interact after that point, then for all
$t>0$
\begin{equation}
G\func{\overline{\Omega_1(0)}}+G\func{\overline{\Omega_2(0)}} \geq
G\func{\overline{\Omega_1(t)}}+G\func{\overline{\Omega_2(t)}}
\label{eq:2ndlaw}
\end{equation}
with equality occurring at time $t$ if, and only if,
$\Omega(t)=\overline{\Omega_1(t)}\otimes \overline{\Omega_2(t)}$

\subsection{Canonical distribution}\label{sss:extremal} The extremal of $G\func{\Omega}$
for a fixed value of $\mean{H}_\Omega=E$ is given by the canonical
distribution:
\begin{equation}
\Omega^{(\beta)}=\frac{e^{-\beta(E) H}}{\trace{e^{-\beta(E) H}}}
\label{eq:canonical}
\end{equation}
where $\beta(E)$ is a parameter depending only on the Hamiltonian
$H$ and the fixed value $E$.

We will assume that the extremal value is always the minimal value,
although this is a far from trivial assumption.

\subsubsection{Subspaces}\label{sss:cansubspace} If there is no interaction term between
subspaces of a canonically distributed system, the marginal
distributions over the subspaces are canonically distributed with
the same $\beta$ parameter.
\begin{eqnarray}
H&=&H_1\otimes I_2 + I_1 \otimes H_2 \\
\Omega^{(\beta)}&=&\frac{e^{-\beta H_1}}{\trace{e^{-\beta
H_1}}}\otimes \frac{e^{-\beta H_2}}{\trace{e^{-\beta
H_2}}}\label{eq:subcanon}
\end{eqnarray}
\subsubsection{Minimising $G+\beta \mean{H}$}\label{sss:canmaxbeta} Given a state space, a
Hamiltonian $H$, the canonical distribution $\Omega^{(\beta)}$ for
that Hamiltonian and any other distribution $\Omega^\prime$ over
that space, then:
\begin{equation}\label{eq:partovi}
G\func{\Omega^\prime}+\beta \mean{H}_{\Omega^\prime} \geq
G\func{\Omega^{(\beta)}}+\beta \mean{H}_{\Omega^{(\beta)}}
\end{equation}
The canonical distribution not only minimises $G$ for a fixed value
of $\mean{H}$ but also minimises $G+\beta\mean{H}$ for a fixed value
of $\beta$.

The result can be rearranged to give
\begin{equation}\label{eq:clausius}
G\func{\Omega^\prime} -G\func{\Omega^{(\beta)}} \geq -\beta \left(
\mean{H}_{\Omega^\prime}-\mean{H}_{\Omega^{(\beta)}} \right)
\end{equation}

\subsubsection{Interactions with arbitrary
distributions}\label{sss:canintarb}
 Interactions between a
canonically distributed system $\Omega_1^{(\beta)}$ and an
arbitrarily distributed system $\Omega_2$.

They are initially $(t=0)$ non-interacting $H=H_1\otimes I_2 + I_1
\otimes H_2$ and uncorrelated
$\Omega(0)=\Omega_1^{(\beta)}(0)\otimes \Omega_2(0)$. The systems
are allowed to interact, $H^\prime=H+V_{12}$, for a finite period of
time, but so that at the end of the interaction
$\mean{H}_\Omega(t)=\mean{H}_\Omega(0)$.

It can then be shown that
\begin{equation}
G\func{\overline{\Omega_2(0)}}+\beta
\mean{H_2}_{\overline{\Omega_2(0)}} \geq
G\func{\overline{\Omega_2(t)}}+\beta
\mean{H_2}_{\overline{\Omega_2(t)}} \label{eq:lucatheorem}
\end{equation}
\subsubsection{Interactions between canonical distributions}\label{sss:canintcan}
Interactions between two canonically distributed systems, but with
different $\beta$ parameters, $\Omega_1^{(\beta_1)}$ and
$\Omega_2^{(\beta_2)}$.

They are initially $(t=0)$ non-interacting $H=H_1\otimes I_2 + I_1
\otimes H_2$ and uncorrelated
$\Omega(0)=\Omega_1^{(\beta_1)}(0)\otimes \Omega_2^{(\beta_2)}(0)$.
The systems are allowed to interact, $H^\prime=H+V_{12}$, for a
finite period of time, outside which
$\mean{V_{12}}_{\Omega(t)}=\mean{V_{12}}_{\Omega(0)}=0$, but at the
end of the interaction $\mean{H}_{\Omega(t)}=\mean{H}_{\Omega(0)}$.

It can be shown that
\begin{equation}
\beta_1\left(\mean{H_1}_{\overline{\Omega_1(t)}}-\mean{H_1}_{\overline{\Omega_1(0)}}\right)
+\beta_2\left(\mean{H_2}_{\overline{\Omega_2(t)}}-\mean{H_2}_{\overline{\Omega_2(0)}}\right)
\geq 0
\end{equation}
Using the notation:
\begin{eqnarray}
\Delta{H_1}&=&\mean{H_1}_{\overline{\Omega_1(t)}}-\mean{H_1}_{\overline{\Omega_1(0)}}\\
\Delta{H_2}&=&\mean{H_2}_{\overline{\Omega_2(t)}}-\mean{H_2}_{\overline{\Omega_2(0)}}\\
\Delta{H}&=& \Delta{H_1}+\Delta{H_2}=0
\end{eqnarray}
this becomes
\begin{equation}
\Delta{H_1}\left(\beta_1 - \beta_2\right)\geq 0
\label{eq:temptheorem}
\end{equation}
\subsection{Large uncorrelated canonical assemblies}\label{ss:luca}
We will now consider a particular type of system called a Large
Uncorrelated Canonical Assembly.

\begin{itemize}
\item The system is large, in that it has a very large number of
degrees of freedom.
\item The distribution over the state space is uncorrelated
with any other system.
\item The distribution over the state space is a canonical distribution,
with parameter $\beta$.
\item The system is an assembly\cite{Per93}.  It consists of a very large
number of identical subsystems, with no interactions between the
subsystems.
\end{itemize}

As the overall distribution is canonical, and there are no
interactions between subsystems, the subsystems have canonical
distributions with the same parameter $\beta$ and will not be
correlated with each other.

\subsubsection{Interactions with arbitrary
distributions}\label{sss:lucainteract} When another system interacts
with a LUCA system, the interaction will always be in a particular
way. The interacting system will have a succession of brief
interactions with successive subsystems of the LUCA, such that no
subsystem of the LUCA is ever encountered twice.

As each interaction with a subsystem is an interaction with a
canonical system with parameter $\beta$, by Equation
\ref{eq:lucatheorem} the value of $G\func{\overline{\Omega_2}}+\beta
\mean{H_2}_{\overline{\Omega_2(t)}}$ for the interacting system will
increase on each interaction.  If there is no further barrier to
prevent it, this value will approach it's maximum.  From the results
of Section \ref{eq:partovi}, the distribution which maximises this
is the canonical distribution over $H_2$ with the parameter $\beta$.

\subsubsection{Interactions with canonical
distributions}\label{sss:lucaquasi} Now consider an interaction
between a LUCA system and a system already canonically distributed
with the same parameter $\beta$.

Consider a very slow variation in the Hamiltonian of the system from
$H_1(0)$ to $H_1(t)$.  We might suppose that we proceed in a series
of small steps.  First isolate the system and make a very small
change in it's Hamiltonian, sufficiently slowly that the quantum
mechanical adiabatic theorem applies\cite{Mes62b}[Ch 17]. As an
isolated system the distribution will move slightly away from
canonical.  Then bring it back into contact with the LUCA, and the
distribution will be restored to a canonical distribution with
parameter $\beta$.

As the steps become infinitesimal, the system remains in the
canonical distribution, but now it is a time varying canonical
distribution:
\begin{equation}
\Omega_1^{(\beta)}(t)=\frac{e^{-\beta H_1(t)}}{\trace{e^{-\beta
H_1(t)}}} \label{eq:isobeta}
\end{equation}
As this is always diagonalised in the $H_1(t)$ basis, we have
\begin{equation}\label{eq:notfree}
\partime{\mean{H_1(t)}_{\overline{\Omega_1(t)}}}=
\mean{\partime{H_1(t)}}_{\overline{\Omega_1(t)}}+ \sum_\alpha
E_\alpha(t)
\partime{ }\left(\frac{e^{-\beta E_\alpha(t)}}{\trace{e^{-\beta
H_1(t)}}}\right)
\end{equation}
where $E_\alpha(t)$ is the instantaneous energy eigenvalue of the
instantaneous eigenstate $\ket{E_\alpha(t)}$ of $H_1(t)$.

Adding the identity
\begin{equation}
\frac{1}{\beta \trace{e^{-\beta H_1(t)}}}\partime{}\trace{e^{-\beta
H_1(t)}}+\frac{1}{\trace{e^{-\beta H_1(t)}}} \sum_\alpha e^{-\beta
E_\alpha(t)}\partime{E_\alpha(t)}=0
\end{equation} the last term can be
rearranged
\begin{eqnarray}
\sum_\alpha E_\alpha(t)
\partime{ }\left(\frac{e^{-\beta E_\alpha(t)}}{\trace{e^{-\beta
H_1(t)}}}\right) &=& -\frac{\beta}{\trace{e^{-\beta
H_1(t)}}}\sum_\alpha E_\alpha(t)e^{-\beta
E_\alpha(t)}\partime{E_\alpha(t)} \eqnwrap
-\frac{1}{\trace{e^{-\beta H_1(t)}}^2} \sum_\alpha
E_\alpha(t)e^{-\beta E_\alpha(t)}\partime{}\trace{e^{-\beta H_1(t)}}
\eqnwrap +\frac{1}{\beta \trace{e^{-\beta
H_1(t)}}}\partime{}\trace{e^{-\beta
H_1(t)}}+\frac{1}{\trace{e^{-\beta H_1(t)}}} \sum_\alpha e^{-\beta
E_\alpha(t)}\partime{E_\alpha(t)} \nonumber \\
&=& \partime{}\left(\frac{\sum_\alpha E_\alpha(t)e^{-\beta
E_\alpha(t)}}{\trace{e^{-\beta H_1(t)}}} +\frac{1}{\beta}\ln
\func{\trace{e^{-\beta H_1(t)}}} \right) \nonumber \nonumber \\
&=&\frac{1}{\beta} \partime{}\sum_\alpha \left( \frac{e^{-\beta
E_\alpha(t)}}{\trace{e^{-\beta H_1(t)}}}\beta E_\alpha(t)+
\frac{e^{-\beta E_\alpha(t)}}{\trace{e^{-\beta H_1(t)}}} \ln
\func{\trace{e^{-\beta H_1(t)}}} \right) \nonumber \\
&=&-\frac{1}{\beta} \partime{}\sum_\alpha \frac{e^{-\beta
E_\alpha(t)}}{\trace{e^{-\beta H_1(t)}}} \ln \func{\frac{e^{-\beta
E_\alpha(t)}}{\trace{e^{-\beta H_1(t)}}} }
=-\frac{1}{\beta}\partime{}G\func{\Omega_1^{(\beta)}(t)}
\end{eqnarray}
Integrating Equation \ref{eq:notfree} from $0$ to $t$ then gives:
\begin{equation}\label{eq:nearlyfree}
D\func{H_1(t)}=\Delta H_1(t)+\frac{1}{\beta}\Delta
G\func{\Omega_1^{(\beta)}}
\end{equation}

\section{Not thermodynamics}
The properties we have considered here are simply mathematical
properties of the Hamiltonian evolution of distributions over state
spaces.  They will apply to any function that has the properties of
being a distribution.  No physical interpretation has been placed
upon them, and no physical interpretation should be placed upon them
unless it can be justified that the property concerned does, in
fact, correspond to a physical property of interest.

We have introduced a particular measure, the Gibbs-von Neumann
measure, which proves to have certain properties. We have also
identified a particular distribution, which is uniquely selected by
that measure. We have identified a particular type of system, which
has that unique distribution.  The resulting description produces
equations that closely resemble thermodynamics.  It is tempting to
identify $G$ with the negative of entropy and see Equation
\ref{eq:2ndlaw} as representing the Second Law of Thermodynamics. It
is tempting to identify the canonical distribution as representing
thermal equilibrium, as it is the state that maximises $-G$, to
identify $\beta$ as the reciprocal of temperature, on the basis of
Equation \ref{eq:temptheorem}, to identify an environmental heat
baths as a LUCA on the basis of Section \ref{sss:lucainteract} and
Equation \ref{eq:nearlyfree} as the isothermal equation $\Delta
W=\Delta E-T\Delta S$. But how justified is all this?

There are two problems.  The first is why should one suppose $-G$
represents entropy or canonical distributions represent thermal
equilibrium?   The second is whether it is even valid to identify
the thermodynamic entropy with a measure upon a distribution.

\subsection{Why \textit{ these} distributions, these measures?}
The first problem can be seen earliest in works such as
\cite{Gib1902}[Chapter XIV], where different distributions are
discussed, which may also appear to reproduce thermodynamic results.
Gibbs cautiously refers only to thermodynamic ``analogies'' in
statistical mechanics (a practice echoed in \cite{Tol1938}, amongst
others).

In his review\cite{Pen1979}, Penrose shows the question remains:
\begin{quote}
what is the physical significance of a Gibbs ensemble? How can we
justify the standard ensembles used in equilibrium theory?
\end{quote}

Let us consider the mathematical structure of the previous section.
The properties derived are almost entirely consequences of two
things:
\begin{enumerate}
\item The function $G$ is a concave function of distributions
(Section \ref{sss:convex});
\item The sum of the marginal values of $G$ for two uncorrelated
distributions is greater that the sum of the marginal values of $G$
for two correlated distributions (Section \ref{sss:gibbssubens}).
\end{enumerate}
If we are tempted to identify $G$ as entropy and $\beta$ as
temperature on the basis of the relationships derived, wouldn't
\textit{ any} non-decreasing, concave function, with the appropriate
property for uncorrelated distributions\footnote{Technically this
property can be obtained for $G$ from (a) concavity; and (b) the
value of $G$ being additive for uncorrelated distributions.  It is
only $G$ that has this property.
 However, this additivity is \textit{ not} necessary for the derived property to
hold, so the derived properties may still hold for other,
non-additive, concave functions.}, be able to do the job?

In recent years it has also been argued that, in quantum mechanics,
it has simply been \textit{ assumed} that the von Neumann entropy is
the appropriate one, and that the only justification offered for it
is flawed:
\begin{quote}
The convention first appears in Von Neumann's \textit{ Mathematical
Foundations of Quantum Mechanics}.  The argument given there to
justify this convention is the only one hitherto offered.  All the
arguments in the field refer to it at one point or another.  Here
this argument is shown to be invalid.\cite{She99a}
\end{quote}

If we assume that the canonical distribution is appropriate for
thermal equilibrium, we may reasonably represent an ideal heat bath
by a LUCA, and from this (for large thermal systems, at least) it is
possible to show that the von Neumann entropy correctly gives the
value of the thermodynamic entropy. But what is the justification
for using the canonical distribution, except that it maximises the
von Neumann entropy?

If we start by identifying a LUCA as an ideal heat bath, we can show
that thermalisation corresponds to approaching the canonical
distribution and so, perhaps, justify the canonical distribution as
appropriate for thermalisation.  But why assume that an ideal heat
bath is a LUCA? A LUCA is canonically distributed already, so
assuming that it represents an environment at some temperature is
tantamount to assuming the very thing we would wish to demonstrate.

If we assume that the von Neumann entropy is the thermodynamic
entropy, then maximising it produces the canonical distribution.
This may justify the canonical distribution as thermal equilibrium
and hence LUCA's as ideal heat baths.  But, without assuming the
canonical distribution is thermal equilibrium in the first place,
what reason do we have for believing the von Neumann entropy is
thermodynamic entropy?

Although we appear to have arrived at expressions that are analogous
to thermodynamic expressions, we cannot identify these expressions
with thermodynamic processes unless we can be sure that they really
are the appropriate representation of the physical process.  There
appears to be a logical gap.

\subsubsection{Why the canonical distribution?}
There has been a large literature devoted to deriving the canonical
distribution.  Attempts in the literature to justify the canonical
distribution are largely to do with the problem of explaining the
approach to thermal equilibrium.  It is assumed that the canonical
distribution is thermal equilibrium (and an entropy usually has
already been decided upon), and the attempt is to explain why
systems are \textit{ in} thermal equilibrium.  While this is not the
same as our concern here, it will be useful to briefly review these
attempts.

The Ehrenfests\cite{EE1912}[Section 25] credit Boltzmann with the
first observation which justifies the canonical ensemble.  The
essence of this justification is that if one takes a large system,
whose distribution is uniform over a constant energy hypersurface
(i.e. a microcanonical distribution), and one takes a small
subsystem of that, then the marginal distribution of the small
subsystem is canonical.  Indeed, with minor variations, this
relationship between the microcanonical and canonical distributions,
is practically the \textit{ only} justification offered in most
textbooks.

The problem then becomes to justify the microcanonical distribution.
Some, following Tolman, simply make a fundamental assumption of a
uniform distribution, with the only justification being, in effect,
the Principle of Insufficient Reason to argue the inappropriateness
any other choice.

Attempts to justify the uniform distribution on dynamical grounds,
argued by the Ehrenfests, have led to the development of the
Boltzmann's ergodic hypothesis, concepts such as metric
transitivity, and weak and strong mixing. Although this has
generated much interesting mathematics, as a justification of the
microcanonical distribution for realistic systems, it can only be
said to have had a mixed degree of success (see \cite{BFK2006} for a
discussion and defense).

A recent development\cite{PSW2005,GLTZ2006} of Boltzmann's original
insight, specific to quantum mechanics, demonstrates that for large
systems in a \textit{pure} quantum state, the reduced density matrix
of a small subsystem is very close to being canonically distributed.
This appears to produce the canonical distribution even without
needing a probability distribution over the whole space.
Unfortunately the result is not true for all pure states, only
``almost every'' or the ``overwhelming majority'' of such states.
The problem here is that these terms are only valid relative to some
measure over the state space and, as it turns out, that measure is
the uniform one. In other words, the development shows that it is
overwhelmingly \textit{ probable} that the individual subsystem
behaves as if it is canonically distributed, \textit{if} we have a
uniform probability distribution over the whole state space.  While
this is a stronger result than Boltzmann's, it cannot be said to
have less problematical assumptions.

\subsubsection{Why the Gibbs-von Neumann measure?}

Once the Gibbs-von Neumann entropy is chosen as physical entropy, it
is possible to argue that the canonical distribution is appropriate
for thermal equilibrium as it maximises
 the entropy of thermally isolated systems.

Our problem here is why the Gibbs-von Neumann entropy
 should be used at all.  This measure
 can certainly be uniquely identified from a number of information theoretic
 prescriptions\cite{Sha48,SW49}.  But why should such information theoretic concerns
should be of any significance for thermodynamics?  Why should
\textit{ thermal} equilibrium have anything to do with maximising
our lack of knowledge?

The idea that entropy is something to do with a lack of knowledge or
uncertainty is an old one, but unless one has already assumed that
the measure of entropy is indeed a function of probability what is
the basis for believing that thermodynamic entropy \textit{ should}
have anything to do with uncertainty?  A priori, what \textit{ is}
the property of thermal states that make us think they represent
maximal ignorance?  Even if this is accepted, there are many
measures of ignorance.  Why are the properties of the Shannon
measure of information the ones that identify the function that
needs maximising?  The existence of alternate information measures,
such as the Renyi measures, and alternate entropy measures, such as
the Tsallis entropy\cite{Tsallis1988,Tsallis2000,Tsallis2003} and
others\cite{BC2007,Campisi2007} call into question whether
assumptions that uniquely specify the Shannon measure can be taken
for granted.

\subsection{Why distributions?  What is entropy anyway?}
The second problem is the complaint of authors such as:
\begin{quote}
The Gibbs entropy is not even an entity of the right sort: It is a
function of a probability distribution, i.e., of an ensemble of
systems, and not a function on phase space, a function of the actual
state $X$ of an individual system\cite{Gol2001}
\end{quote}
\begin{quote}
thermodynamic entropy is patently an attribute of \textit{
individual systems}. And attributes of individual systems can
patently be nothing other than attributes of the \textit{ individual
microconditions}.\cite{Alb01}\end{quote}
\begin{quote}
for present purposes - reconciling thermodynamics with mechanics -
[Gibbs entropy] is of no use since thermodynamic entropy is
applicable to individual systems.  My coffee in the thermos has an
objective thermodynamic entropy as a property.\cite{Cal2004}
\end{quote}

The literature abounds with alternative definitions of entropy (in a
recent work\cite{CS2005}[Chapter 1] 21 different versions of entropy
are listed).  Perhaps the question should not be ``What is the
correct expression for entropy?'', but ``What exactly is `entropy'
supposed to \textit{ be}?''.  What is it about a particular
expression that legitimates referring to it as `entropy'?  What, in
short, is `entropy' \textit{for}?  To develop a physical
understanding of this, we will let the answer emerge \textit{ from}
statistical mechanics, rather than be presupposed.

\section{Statistical Mechanics}\label{s:probabilities}
After having asked many questions in the previous Section, we will
now proceed by ignoring them.  We will develop statistical mechanics
without any reference to entropy at all.  The main purpose of this
is to demonstrate that much of the physical understanding of
statistical mechanics may be developed without any reference to
thermodynamics.

The basic assumption of this section is that we have a physical
system of interest where it is valid to talk about a probability of
the system being in a particular state.  We will not consider why
such a probabilistic situation has occurred, and will attempt to
avoid all discussion of what `probability' actually means.  Instead
we will take for granted that we are dealing with situations where
statements of the form ``There is a probability $p(X)$ that the
system is in state $X$'' are meaningful, and work through the
consequences of this.  From now on, when we refer to a system, we
will mean a physical system, with a state space, a Hamiltonian
evolution and probability for the system being in any particular
state.

\subsection{Operators and evolutions}
To go back to basics, we start by deciding what we can say about the
average value of observing an observable.  Suppose we have an
observable $A$, then the expectation value of the observable, when
the system is in state $\ket{\alpha_n}$ is
$\ip{\alpha_n}{A}{\alpha_n}$. If the state $\ket{\alpha_n}$ has
probability $p(\alpha_n)$, then expectation value of the observable
is
\begin{equation}
\mean{A}=\sum_n p(\alpha_n) \ip{\alpha_n}{A}{\alpha_n}
\end{equation}
This can be rewritten as
\begin{equation}
\mean{A}=\trace{\rho A}
\end{equation}
where
\begin{equation}
\rho=\sum_n p(\alpha_n) \proj{\alpha_n}
\end{equation}
(Note we have not assumed that the set $\basis{\alpha_n}$ are an
orthonormal basis).

$\rho$ is the density matrix for the system.  All the statistical
properties of the system can be calculated from the density matrix.
As well as mean values we may also calculate variances, standard
deviations, and indeed all of the standard apparatus of statistics
and probability theory.

We also note that the density matrix fulfils the criteria for a
distribution, provided that (for isolated systems, at least), if the
state $\ket{\alpha_n}$ at time $t=0$, has probability $p(\alpha_n)$,
the state evolves into $\ket{\alpha^\prime_n}$ at a later time
$t=\tau$, by a Hamiltonian evolution and state
$\ket{\alpha^\prime_n}$ has probability $p(\alpha_n)$.

At the risk of further stating the obvious, let us just remember a
few other things. The expectation value is not telling us the exact
value that we will actually get, nor is it even telling us that the
value we will actually get is close to this value.  We should no
more expect this than expect that when we roll a die, the face
should come up with a number close to three and a half.

The value we get is an expectation value, because that is the
statistical property we have chosen to calculate.  Statistics is
certainly not limited to calculating expectation values!  If we want
to calculate other statistical properties, perform other statistical
operations, the density matrix certainly allows us to do so.  If we
find physical reasons for preferring other statistics, then those
other statistics are what we should use.  There is nothing
intrinsically special to expectation values!  Finally, to avoid
cumbersome words, from now on we will refer to the expectation value
of properties as the mean value.

\subsection{Isolated systems}
If, as is usual, we identify the Hamiltonian as the energy operator,
then the mean energy of the system is
\begin{equation}
\mean{H}_\rho=\trace{H\rho}
\end{equation}

We now consider how this mean value varies with time.

\subsubsection{Work}
For an isolated\footnote{Note that we do not take isolated to mean
having a time independant Hamiltonian.  We take isolated to mean
only that there is no interaction Hamiltonian with another system.}
system, the density matrix evolves unitarily:
\begin{equation}
i \hbar\partime{\rho}=\com{H}{\rho}
\end{equation}
so the mean energy changes:
\begin{equation}
\partime{\mean{H}_\rho}=\mean{\partime{H}}_\rho
\end{equation}
Still back to basics, we might ask what this means.  The left hand
side is clearly the rate at which the mean energy of the system is
changing. What of the right?

Let us suppose that the Hamiltonian is a function of some parameters
$(x,y,z)$ that are varying in time:
\begin{equation}
H=\sum_n E_n(x,y,z)\proj{E_n(x,y,z)}
\end{equation}
The eigenstates can be rewritten
\begin{equation}
\ket{E_n(x,y,z)}=\Upsilon(x,y,z)\ket{E_n}
\end{equation}
so that the operator
\begin{equation}
\partime{H}=\sum_n \underline{\dot{x}}\cdot \nabla E_n(x,y,z)\proj{E_n(x,y,z)}
+\com{\underline{\dot{x}}\cdot \underline{\Theta}}{H}
\label{eq:genwork}
\end{equation}
where
\begin{eqnarray}
\underline{\dot{x}}&=&\left(\partime{x},\partime{y},\partime{z}\right)\\
\underline{\Theta}&=&\left(\partxy{\Upsilon(x,y,z)}{x}\Upsilon^\dag(x,y,z),
\partxy{\Upsilon(x,y,z)}{y}\Upsilon^\dag(x,y,z),\partxy{\Upsilon(x,y,z)}{z}\Upsilon^\dag(x,y,z)\right)
\end{eqnarray}
The first part of Equation \ref{eq:genwork} should be recognised as
the generalised force $\nabla E_n(x,y,z)$ that comes from the change
in energy eigenvalues due to a change in the parameters $(x,y,z)$.
The product with the rate of change of those parameters gives the
rate of work against the force.

The second part is slightly more subtle.  If the energy eigenstates
are varying, then, if we were to keep the energy eigenvalues and the
state of the system fixed, the expectation value of the energy for
that state would be changing.  This term, therefore, represents the
rate of work required to rotate the energy eigenstates.

The term $\ip{\alpha_n}{\partime{H}}{\alpha_n}$ gives the mean rate
of work required for state $\ket{\alpha_n}$, so
$\mean{\partime{H}}_\rho$ is just the mean rate of work required,
given the density matrix $\rho$.

So, the rate at which the mean energy of the system is changing is
equal to the mean rate at which work is being performed upon the
system. Given the systems is isolated, we should hope so!

\subsubsection{Adiabatic availability} We now ask the question: how
much work may be extracted from a given state $\rho$, by a cyclic
variation of the system Hamiltonian\footnote{Note, here, that this
is a cyclic variation in the Hamiltonian, \textit{not} the state of
the system.  This is not, therefore, directly related to the Kelvin
statement of the second law.}
 $H$?  We will call this the adiabatic availability of the state
 $\rho$ (see \cite{GB1991}[Chapter 5]).
We assume that the system is completely isolated.   Clearly the work
extracted by a variation in the Hamiltonian is given by
\begin{equation}
W=-\int_0^\tau \partime{\mean{H}_\rho} dt
\end{equation}
As the system is isolated, the evolution of $\rho$ depends solely
upon $H$, and we require that $H=H_0$ for all $t<0$ and $t>\tau$.
This allows us to rewrite the result, rather trivially, as:
\begin{equation}\label{eq:workoutput}
W=-\trace{H_0 U^\dag(\tau) \rho_0 U(\tau)-H_0 \rho_0}
\end{equation}
where
\begin{equation}
\rho_0 =\sum_n \lambda_n \proj{n}
\end{equation} is the initial
density matrix,
\begin{equation}
i \hbar \partime{U(t)}=H(t) U(t)
\end{equation}
and $U(0)=I$.

We would clearly like to extract as much work as possible from the
system.  Is there a limit to how much we can extract?  If the
initial Hamiltonian $H_0$ is not bounded from below (i.e. does not
have a ground state with finite energy) then the answer is, no,
there is no limit.  Excluding this case, the answer is, yes.

\subsubsection{Passive distributions} Let us note that Equation
\ref{eq:workoutput} expresses the difference between the mean energy
of the initial density matrix $\rho_0$, and another density matrix
which can be related to it by a unitary transformation $U(\tau)$. To
get the most work out, it is therefore necessary to vary the
Hamiltonian so as minimise the mean energy of the state
$\rho(\tau)=U^\dag(\tau) \rho_0 U(\tau)$.  As $\rho(\tau)$ must have
the same eigenvalues as $\rho_0$, it turns out that the minimum is
just the state which is diagonalised in the energy eigenstates
\begin{equation}
\rho(\tau)=\sum_n p_n \proj{E_n}
\end{equation}
such that
\begin{equation}\label{eq:passivity}
p_m \geq p_n \Leftrightarrow E_m \leq E_n
\end{equation}
A density matrix which satisfies these criteria is called a passive
distribution.  Intuitively it is clear that a passive distribution
must certainly minimise the internal energy for a set of
eigenvalues, if diagonalised in the energy eigenstates.  If Equation
\ref{eq:passivity} did not hold, for two given states, it would
always be possible to reduce the mean energy (and extract work) by
swapping those two states.

We now show that density matrices that are not diagonalised by the
energy eigenstates are not passive.  Take a density matrix which is
assumed to not be diagonalised in the energy basis:
\begin{equation}
\rho_1 = \sum_j \lambda_j \proj{\Lambda_j}
\end{equation}
with the ordering $\lambda_i \geq \lambda_j \Leftrightarrow i \leq
j$, and form the density matrix diagonalised in the energy basis:
\begin{equation}
\rho_2 = \sum_i \mu_i \proj{E_i}= \sum_i \ev{E_i}{\rho_1} \proj{E_i}
\end{equation}
for which clearly $\trace{H\rho_1}=\trace{H\rho_2}$.  Now compare
$\rho_2$ with the passive distribution
\begin{equation}
\rho_3 = \sum_j \lambda_j \proj{E_j}
\end{equation}
As the eigenvalues are related by a doubly stochastic map
\begin{equation}
\mu_i=\sum_j \magn{\bk{E_i}{\Lambda_j}} \lambda_j
\end{equation}
it follows\cite{HLP1934} that
\begin{equation}
\sum_i \mu_i E_i \geq \sum_j \lambda_j E_j
\end{equation}
with equality possible if and only if the doubly stochastic map is a
permutation.  For the $\mu_i$ distribution to be passive, this would
require an identity map, which is not the case by assumption. It
follows that $\trace{H\rho_1} > \trace{H\rho_3}$, but $\rho_3$ is
clearly accessible from $\rho_1$ by a unitary map.  $\rho_1$ is
therefore not passive.

For any given density matrix $\rho$ and Hamiltonian $H_0$, there is
a passive distribution $\widetilde{\rho}$ with the same eigenvalues.
The adiabatic availability is then:
\begin{equation}
A\func{\rho,H_0}=\trace{H_0 \left(\rho- \widetilde{\rho}\right)}
\geq 0
\end{equation}
Equality is reached only for passive distributions, which all have
adiabatic availabilities of zero.  The adiabatic availability is
always uniquely defined although the passive distribution is unique
only if the energy eigenvalues are non-degenerate.

For an isolated system it is clearly the case that the work
performed upon the system, in any cyclic variation of the
Hamiltonian, must equal the change in adiabatic availability. For a
non-cyclic variation in the Hamiltonian, this is not the case.  If
$H_0 \rightarrow H_1$ leads to $\rho_0 \rightarrow \rho_1$ then the
work $W$ and change in adiabatic availability $\Delta A$ are related
by
\begin{eqnarray}
W &=& \trace{H_1 \rho_1}-\trace{H_0 \rho_0} \\
\Delta A &=& W - \trace{(H_1-H_0)\widetilde{\rho_0}}
\end{eqnarray}

An explicit cyclic Hamiltonian capable of extracting the available
energy is given (for $0<t<\tau$) by:
\begin{equation}
H(t)=H_0 \cos\left(\frac{2 \pi t }{\tau}\right)-\frac{2 \imath
\hbar}{\tau}\sin^2\left(\frac{\pi t }{\tau}\right)\ln \left[ \sum_j
\kb{E_j}{\Lambda_j} \right]
\end{equation}

We will note the following in passing: that if the energy spectrum
is bounded from above, then there is also \textit{maximum} amount of
work that may be unitarily performed \textit{upon} the system, and a
set of distributions for which that maximum is zero. These
distributions have the property:
\begin{equation}\label{eq:passive}
p_m \leq p_n \Leftrightarrow E_m \leq E_n
\end{equation}
Unless otherwise stated, we will assume that the energy spectrum is
not bounded from above, in which case these distributions will not
have finite mean energy, and we will not consider them.

\subsection{Interacting systems} We now move on to the situation
where we have two systems that are allowed to interact for a period
of time, so that $H(t)=H_1\otimes I_2 +I_1 \otimes H_2+V_{12}$.  The
combined system is isolated, except through the variation of the
Hamiltonian.

First we have
\begin{eqnarray}
\partime{\mean{H}_\rho}&=&\partime{\mean{H_1}_{\overline{\rho_1}}}
+\partime{\mean{H_2}_{\overline{\rho_2}}}
 +\partime{\mean{V_{12}}_{\rho}} \\
&=&\mean{\partime{H}}_{\rho} \\
&=&\mean{\partime{H_1}}_{\overline{\rho_1}}
+\mean{\partime{H_2}}_{\overline{\rho_2}}
+\mean{\partime{V_{12}}}_{\rho}
\end{eqnarray}
with the middle line because the combined system is not interacting
with any third system.  We should not be surprised to see that the
rate at which the mean energy of the combined system changes, is
equal to the mean of the rate at which work is performed upon the
subsystems and the interaction between them.

We also have
\begin{eqnarray}
i \hbar
\partime{\mean{H_1}_{\overline{\rho_1}}}&=& i
\hbar\mean{\partime{H_1}}_{\overline{\rho_1}}+ Q\func{H_1}\\
i \hbar
\partime{\mean{H_2}_{\overline{\rho_2}}}&=& i
\hbar\mean{\partime{H_2}}_{\overline{\rho_1}}+ Q\func{H_2}\\
i \hbar
\partime{\mean{V_{12}}_{\rho}}&=& i
\hbar\mean{\partime{V_{12}}}_{\rho}-Q\func{H_1}-Q\func{H_2}
\end{eqnarray}
where for convenience we define
\begin{eqnarray}
Q\func{H_1}=&\mean{\com{H_1}{V_{12}}}_{\rho}&=\sum_n i \hbar
\partime{p(\alpha_n)}\ev{\alpha_n}{H_1}+
\mean{\com{H_1}{\Theta_1}}_{\overline{\rho_1}}  \\
Q\func{H_2}=&\mean{\com{H_2}{V_{12}}}_{\rho}&=\sum_n i \hbar
\partime{p(\beta_n)}\ev{\beta_n}{H_2}+
\mean{\com{H_2}{\Theta_2}}_{\overline{\rho_2}}  \\
\end{eqnarray} with
\begin{eqnarray}
\overline{\rho_1}&=&\sum_n p(\alpha_n,t)\proj{\alpha_n(t)}=\partrace{\rho(t)}{2}\\
\overline{\rho_2}&=&\sum_n
p(\beta_n,t)\proj{\beta_n(t)}=\partrace{\rho(t)}{1}
\end{eqnarray} and
$\Theta_1$ and $\Theta_2$ are defined, as in Section
\ref{ss:operators}, as the Hamiltonian operators
\begin{eqnarray}
\ket{\alpha_n(t)}&=&\Upsilon_{\alpha}(t)\ket{\alpha_n(0)} \\
i \hbar \partime{\Upsilon_{\alpha}(t)}&=&\Theta_1(t) \Upsilon_{\alpha}(t) \\
\ket{\beta_n(t)}&=&\Upsilon_{\beta}(t)\ket{\beta_n(0)} \\
i \hbar \partime{\Upsilon_{\beta}(t)}&=&\Theta_2(t) \Upsilon_{\beta}(t) \\
\end{eqnarray}

The term $Q\func{H_1}$ clearly represents the mean rate at which
energy is flowing into system 1, in addition to work being performed
upon it, and similarly for $Q\func{H_2}$ and system 2. Remember also
that if any two of $H_1$, $\Theta_1$ and $\overline{\rho_1}$
commute, the commutator term is zero.

We will now consider some simplifying conditions.

\begin{enumerate}
\item{Constant interaction potential}

 If the only work being performed upon the joint system is
through $H_1$ and $H_2$, then $\partime{V_{12}}=0$:
\begin{equation}
i \hbar
\partime{\mean{V_{12}}_{\overline{\rho}}}+
Q\func{H_1}+ Q\func{H_2}=0
\end{equation}

We now use the notation
\begin{eqnarray}
\Delta X(t)&=&\int_0^{t}
\partime{\mean{X}_{\Omega}} dt =\mean{X(t)}_{\Omega(t)}-\mean{X(0)}_{\Omega(0)} \\
D\func{X(t)}&=&\int_0^{t} \mean{\partime{X(t)}}_{\Omega(t)} dt
\end{eqnarray}
to consider how the mean energy of the systems change after a finite
period of interaction.  $\Delta H$ gives the change in the mean
energy of the system over the course of the interaction, and
$D\func{H}$ gives the mean work performed upon the system.

If the interaction term is constant in time, then
\begin{equation}
\Delta V_{12}+\int_0^t Q\func{H_1}dt + \int_0^t Q\func{H_2} dt =0
\end{equation}

\item{Finite interaction duration}

We now suppose that the systems are initially separated so that
$\trace{V_{12}(0)\rho(0)} = 0$ and that at time $t$ they have been
separated\footnote{The moving of the systems can be achieved by
variations in the internal Hamiltonians of each, so this does not
conflict with $\partime{V_{12}}=0$.} again $\trace{V_{12}(t)\rho(t)}
= 0$.
\begin{eqnarray}
\Delta Q&=&\int_0^t Q\func{H_1}dt =-\int_0^t Q\func{H_2} dt\\
\Delta H_1&=&D\func{H_1}+\Delta Q \\
\Delta H_2&=&D\func{H_2}-\Delta Q
\end{eqnarray}

$\Delta Q$ is the mean flow of energy between the two systems,
during the interaction.

\item{No change to second system}

Finally we consider the effect of $\partime{H_2}=0$ :
\begin{eqnarray}
D\func{H_1}&=&\Delta H_1-\Delta Q \\
\Delta H_2&=&-\Delta Q
\end{eqnarray}

The interaction with the second system can allow energy to flow from
system 2 into system 1.  If the work done on system 1 is negative,
$D\func{H_1}<0$, the energy flow can be extracted. This can still be
true if the variation in $H_1$ is cyclic, and even if $\Delta H_1
=0$.
\end{enumerate}

\subsubsection{Completely passive distributions} We complete the
notion of adiabatic availability by noting the need for a stronger
notion than passivity is required if composite systems are
considered.  Let us consider a joint system, consisting of a joint
Hamiltonian:
\begin{equation}
H_{12}=H_1 \otimes I_2 + I_1 \otimes H_2
\end{equation}
and a joint density matrix
\begin{equation}
\rho_{12}=\rho_1 \otimes \rho_2
\end{equation}
such that
\begin{eqnarray}
H_1 &=& \sum_n E_n \proj{E_n^{(1)}} \\
H_2 &=& \sum_n E_n \proj{E_n^{(2)}} \\
\rho_1 &=& \sum_n p_n \proj{E_n^{(1)}} \\
\rho_2 &=& \sum_n p_n \proj{E_n^{(2)}}
\end{eqnarray}
Even if $\rho_1$ is a passive distribution (and, by construction, so
will $\rho_2$ be), the combined density matrix $\rho_{12}$ may not
be a passive distribution.

To show this, we need only consider three energy levels, and their
probabilities:
\begin{eqnarray}
E_1 < & E_2 & < E_3 \\
p_1 > & p_2 & > p_3
\end{eqnarray}
For the joint system to be passive it is necessary that
\begin{equation}
(2 E_2 -E_1 -E_3)((p_2)^2-p_1p_3) < 0
\end{equation}
It is a simple matter to find values\footnote{Consider
\begin{eqnarray}
E_1=1 & E_2 =3 & E_3=4 \nonumber \\
p_1=1/2 & p_2=1/3 & p_3=1/6 \nonumber
\end{eqnarray}
Energy may be extracted by swapping the $\ket{E_1 E_3}$ and the
$\ket{E_2 E_2}$ states.} for which this fails and equally easy to
find values\footnote{$E_2=2$} for which this holds.

If a product of 2 equivalent passive systems, is passive, then the
distribution may be termed 2-passive. Similarly, if a product of $N$
equivalent passive systems is passive, then the distribution may be
termed $N$-passive.  A completely passive distribution (equivalent
to the mutual stable equilibrium of \cite{GH1976b}) is one that is
$N$-passive, for all finite $N$.

The necessary and sufficient condition for $N$-passivity is that,
for all combinations of natural numbers $\{a_i\}$ and $\{b_j\}$ such
that
\begin{equation}
\sum_i a_i=\sum_j b_j=N
\end{equation}
then
\begin{equation}
\left(\sum_i a_i E_i \leq \sum_j b_j E_j\right) \Longleftrightarrow
\left(\prod_i {p_i}^{a_i} \geq \prod_j {p_j}^{b_j}\right)
\end{equation}

To simplify this, consider just three levels, $E_i < E_j < E_k$, and
$b_j=N$. It is necessary that either
\begin{eqnarray}
N E_j &\leq& (N-n) E_i - n E_k \nonumber \\
(p_j)^N &\geq & (p_i)^{(N-n)}(p_k)^n
\end{eqnarray}
or
\begin{eqnarray}
N E_j &\geq &(N-n) E_i - n E_k \nonumber \\
(p_j)^N & \leq & (p_i)^{(N-n)}(p_k)^n
\end{eqnarray} for $0 < n < N$.

Now as $N E_i < N E_j < N E_k$ and $(p_i)^N
> (p_j)^N
> (p_k)^N$, there must exist real numbers $0<l_{(ijk)},m_{(ijk)}<N$ such that
\begin{eqnarray}
N E_j &=& (N-l_{(ijk)}) E_i - l_{(ijk)} E_k \nonumber \\
(p_j)^N &=& (p_i)^{(N-m_{(ijk)})}(p_k)^{m_{(ijk)}}
\end{eqnarray}
For the system to be $N$-passive, $l_{(ijk)}$ and $m_{(ijk)}$ cannot
be separated by any integers (as any such integer will yield an $n$
for which the $N$-passive conditions fail). This must hold for all
$(ijk)$ triples of energy levels.

If we rewrite the above equations as
\begin{equation}
l_{(ijk)}=N\frac{E_j-E_i}{E_k-E_i}
\end{equation}
\begin{equation}
m_{(ijk)}=N\frac{\ln(p_i)-\ln(p_j)}{\ln(p_i)-\ln(p_k)}
\end{equation}
then
\begin{equation}
l_{(ijk)}-m_{(ijk)}=N\left(\frac{E_j-E_i}{E_k-E_i}-\frac{\ln(p_i)-\ln(p_j)}{\ln(p_i)-\ln(p_k)}\right)
\end{equation}
Unless
\begin{equation}
\frac{E_j-E_i}{E_k-E_i}=\frac{\ln(p_i)-\ln(p_j)}{\ln(p_i)-\ln(p_k)}
\end{equation}
it is clear that if $N$ becomes sufficiently large then
$\mod{l_{(ijk)}-m_{(ijk)}}>1$.  When this happens the combined
distribution is no longer passive, as there must exist an integer
$n$ between $l_{(ijk)}$ and $m_{(ijk)}$.

To be completely passive, it must be the case that for each triple
$l_{(ijk)}=m_{(ijk)}$.  This leads to
\begin{equation}
\frac{\ln(p_i/p_j)}{E_j-E_i}=\frac{\ln(p_i/p_k)}{E_k-E_i}
\end{equation}
For this to hold for any triplet $(ijk)$ of energy levels, then
\begin{equation}
\frac{\ln(p_i/p_j)}{E_j-E_i}=\beta
\end{equation}
where $\beta$ is a constant.  Further rearranging shows
\begin{equation}
\ln(p_i)+\beta E_i= \ln(p_j)+\beta E_j=\ln \lambda
\end{equation}
where $\lambda$ is also a constant, giving\footnote{Note: passivity
requires $\beta >0$.}:
\begin{equation}
p_i=\lambda e^{- \beta E_i}
\end{equation}
for all $i$.  It is a simple matter to verify this distribution is
sufficient for $N$-passivity.  As the value of $N$ has disappeared,
it is apparent that the canonical distribution must be $N$-passive,
for all any $N$.  It follows that the canonical distribution is the
unique, completely passive distribution, for separable Hilbert
spaces\cite{PW1978,Lenard1978,Sewell1980}.

\subsection{Summary}
Does much of the above seem somewhat obvious?  We should hope so!
All we have done is to apply the normal rules of probability theory
to Hamiltonian evolutions.  We have derived average terms for the
rate of change of energy, rate of work and the flow of energy
between two interacting systems.  We need make no reference to
thermodynamic concepts to do this.  There is no need to introduce
approximations and the results apply to systems with any (finite)
number of degrees of freedom.

Remarkably, we have shown how there is maximum amount of energy, the
adiabatic availability of the system, that can be extracted as work
from an isolated system, in a specific type of cyclic process.  Not
all of the energy of a system is available for work. This conclusion
can be drawn without needing to invoke notions of entropy or
consider thermal heat baths or engines operating between them.

Most remarkably we have a concept of passivity, that seems similar
to thermal equilibrium, and a stronger concept of complete
passivity, or mutual stable equilibrium, and the only completely
passive distribution is the canonical distribution! Yet we have at
no point referred to any thermal concepts, whether entropy,
temperature or thermal equilibrium.

There may be disagreement over when a probabilistic statement can be
justified.  There may be disagreement over what such a statement
means.  There may even be disagreement over what the value of the
probability is.  Whenever probabilistic statements are justified,
the results given here follow.

\section{Statistical Temperature} \label{s:temperature}
Having established the rules of statistical mechanics, we now need
to see if it can account for thermal phenomena.  We wish to avoid,
as far as possible, assuming any of the structure of thermodynamics,
and instead focus upon the physical phenomena.  It is tempting to go
from complete passivity to a Large Completely Passive Assembly, and
then identify this with an environmental heat bath.  We will
continue to resist temptation!  We still have not yet justified that
the physical systems we characterise as thermal states are actually
canonical distributions.

To do this, we will deduce, from a set of observations, that there
is only one possible way to represent thermal states in the context
of statistical mechanics and that is the canonical distribution.  No
reference to ideas of entropy, heat baths, approaches to
equilibrium, or information theory, will need to be used.  The
analysis here was inspired largely by \cite{Szi1925}, although the
presentation differs quite significantly.

\subsection{Some properties of temperature}

Statistical mechanics has a far broader scope than thermal
phenomena. To see how statistical mechanics deals with thermal
phenomena we must first identify what thermal phenomena are, and how
this restricts the description of the systems to which we wish to
apply the methods of statistical mechanics.  We do not derive the
concept of temperature.  Instead we consider temperature an
empirically observed phenomena and ask what the theoretical
description of such phenomena could be.  What are the empirical
properties we know about these thermal systems, and what constraints
does this lay upon what physical states can represent them?

Our approach will be to start by examining the phenomena of
temperature, taking it for granted that we have some notion of
temperature from our experience of things being \textit{ hot} and
\textit{ cold}.  We will not need to consider what it means for one
system to be \textit{ hotter} than another, only what it means for
two systems to be at the same temperature as each other.

We will state the following properties of two systems that are at
the same temperature:

\begin{enumerate}
\item{No spontaneous flow of energy.}

If two systems in isolation, are at the same temperature, then if
they interact with each other there can be no \textit{mean} flow of
energy between them.  Energy may be exchanged in individual systems,
as fluctuations, but the expectation value for the exchange must be
zero.

\item{Composition.}

Temperature is not changed by combining systems at the same
temperature.  When two systems are each individually at the same
temperature as each other, then the joint system that is formed by
combining those two systems, is a system at the same temperature.

\end{enumerate}

These two observations are all we need to derive thermal physics
from statistical mechanics.  We will also show that the composition
property may be replaced by the following two conditions:
\begin{enumerate} \setcounter{enumi}{1}
\item{}
\begin{enumerate}

\item{Transitivity.}

Temperature is transitive between systems. When two systems are each
individually at the same temperature as a third system, then they
are at the same temperature as each other.

\item{Universality.}

The property of being at a \textit{particular} finite temperature,
can hold for every possible system.  So, for any given Hamiltonian,
there exists at least one distribution that corresponds to each
temperature.

\end{enumerate}
\end{enumerate}
\subsection{Deriving the temperature distribution}
We wish to regard the previous statements as providing a set of
empirical observations that we are going to use to deduce how
thermal states needs to be treated in statistical mechanics.

\begin{enumerate}
\item{No spontaneous flow of energy.}

First we must clarify what is the observed phenomena we are
proposing, and secondly, how that can be represented.

In general, when two objects, initially non-interacting, are brought
into contact, and allowed to interact through that contact, when
they are separated their states have changed.  Careful calorimetry
experiments, based upon the work required to effect the equivalent
changes to those systems, when isolated from each other, allows us
to identify a quantity of energy that was exchanged between the two
systems.

When systems are at the same temperature, the quantity of energy
exchanged is zero.  This needs two qualifications.  When the
measurements are sufficiently sophisticated to include fluctuation
phenomena, the energy exchanged in any one instance, of contact
between two systems, may be non-zero. The energy exchange is still
zero \textit{on average}.  Also, we must exclude interactions that
can cause chemical or nuclear reactions.  To do this we require the
output density matrices to have non-zero eigenvalues only in those
regions of state space for which the input density matrices have
non-zero eigenvalues.

Before contact, the systems are separated, and so described by a
product density matrix with a non-interacting Hamiltonian:
\begin{eqnarray}
\rho_0 &=&\rho_1 \otimes \rho_2 \nonumber \\
H_0 &=& H_1 \otimes I_2 + I_1 \otimes H_2
\end{eqnarray}
In principle we could allow an interaction Hamiltonian $V_{12}$
subject to $\trace{V_{12}\rho_0}=0$.

The two systems are now brought into contact.  This can happen in
two possible ways: the interaction Hamiltonian $V_{12}$ is changed,
so it becomes non-zero for $\rho_0$; or at least one of the system
Hamiltonians is changed so that $\rho_0$ itself evolves into a state
$\rho$ for which $\trace{V_{12}\rho} \neq 0$.  We will represent
both cases by simply assuming overall a time varying Hamiltonian,
$H(t)$, such that $H(t)=H_0 \forall t<0$.  As the interaction is of
a time limited duration, we also require there exist a time $\tau$
for which $H(t)=H_0 \forall t>\tau$.

We now want to focus on the concept of a \textit{spontaneous} flow.
We cannot take this to mean in the absence of all interventions, as
we are having to intervene to bring the systems into contact, then
separate them.  This involves a time variation in the Hamiltonian,
and the possibility of work being performed upon the system.  For a
flow of energy to be spontaneous, therefore, we add the restriction
that the net work performed upon the joint system, over the course
of the interaction, be zero.

We can summarise this as follows:
\begin{eqnarray}
U&=&e^{\imath \int_0^\tau H(t)} \nonumber \\
\rho_\tau &=& U\rho_0 U^\dag \nonumber \\
W &=& \trace{H_0 \left(\rho_\tau-\rho_0\right)} \nonumber \\
\Delta E_1 &=& \trace{H_1 \otimes I_2 \left(\rho_\tau-\rho_0\right)}
\nonumber \\
\Delta E_2 &=& \trace{I_1 \otimes H_2
\left(\rho_\tau-\rho_0\right)}= W- \Delta E_1
\end{eqnarray}
A necessary condition for $\rho_1$ and $\rho_2$ be at the same
temperature as each other is that, for all $H(t)$ such that $W=0$,
then $\Delta E_1=0$.  In other words, if the two systems are at the
same temperature, then provided no work is performed upon the joint
system, there can be no exchange of energy between the two systems.

We can also see that, for this condition to be met, it is necessary
to consider only the overall unitary evolutions $U$, and not the
detailed interaction $H(t)$.  If we rewrite the condition as: for
$\rho_1$ and $\rho_2$ be at the same temperature as each other, then
for all $U$ for which $W=0$, it must be the case that $\Delta
E_1=0$; we can immediately note that the system $\rho_0$ must be a
passive system.

If $\rho_0$ is not a passive system, then it is possible to extract
its adiabatic availability as work.  We can then apply that same
quantity of work, to either system, in isolation.  The net work is
zero, but energy can be exchanged.  It follows that if $\rho_0$ is
not passive, there exists a unitary evolution that violates the
conditions for $\rho_1$ and $\rho_2$ to be at the same temperature,
as a Hamiltonian interaction may always be constructed to implement
the unitary evolution.

\item{Composition.}

If two systems are individually at the same temperature $T$, then
the combined system is also at that same temperature $T$.  This is a
surprising property.  It is not directly deducible from the more
familiar transitivity property of temperature, but certainly
embodies one of our intuitive notions of what it means for two
systems to be at the same temperature.

Let us express the concept of `being at the same temperature' as a
relationship `$\sim$', so that $\rho_1 \sim \rho_2$ means $\rho_1$
and $\rho_2$ are at the same temperature\footnote{The relationship
is clearly symmetric, so that $\rho_1 \sim \rho_2 \Leftrightarrow
\rho_2 \sim \rho_1$ and reflexive, so that $\rho_1 \sim \rho_1$.}.
The compositional property states:

\begin{equation}
\rho_1 \sim \rho_2 \Rightarrow \rho_1 \sim \rho_1 \otimes \rho_2
\end{equation}

We also know, from the first property, that if $\rho_1 \sim \rho_2$,
then $\rho_1 \otimes \rho_2$ is passive.  As $\rho_1 \sim \rho_1$,
from induction it follows that $\rho_1 \otimes^N \rho_1$ is passive,
for all $N$.  It follows for $\rho_1$ to be a temperature state,
that it must be completely passive, and hence a canonical
distribution.

\end{enumerate}

This compositional property of thermal states is rarely emphasised,
although it plays a very strong role in our intuitive sense of what
is required of a state, for it to be thermal.  The transitive
property of temperature is more usually encountered.  Can
transitivity be used to deduce the canonical distribution?

\begin{enumerate} \setcounter{enumi}{1}
\item{}
\begin{enumerate}

\item{Transitivity}

The transitivity of temperature is closely related to the
operational requirement that any two systems, when usable as
thermometers, must agree when systems are at the same temperature.
When applied to statistical states, it defines a necessary, but not
sufficient, requirement for the states to be considered at the same
temperature.

The first requirement gives us a necessary condition for $\rho_1$
and $\rho_2$ to be at the same temperature: that $\rho_1 \otimes
\rho_2$ be passive.  Suppose there is a third system $\rho_3$.  The
fact that $\rho_1 \otimes \rho_3$ and $\rho_2 \otimes \rho_3$ may
both be passive is not sufficient to ensure $\rho_1 \otimes \rho_2$
is passive. Transitivity is, therefore, a further restriction.

However, if $\rho_1 \otimes \rho_2$ is not passive, we are unable to
say whether it is $\rho_1$ or $\rho_2$ (or neither) that could still
be regarded as being at the same temperature as $\rho_3$.  All that
we can say is that, for a collection of distributions, $\{\rho_1,
\ldots, \rho_n\}$, to be all at the same temperature, $T$, it is
necessary that every combination $\rho_i \otimes \rho_j$ be a
passive distribution.  If we introduce a new distribution, which is
only jointly passive with \textit{some} of the original collection,
is it the new system that is not at the same temperature, or was it
our original collection, that was not a true collection of
distributions at the same temperature?

\item{Universality}
We now introduce the assumption that temperature be
\textit{universal}. By this we mean that, for any system
Hamiltonian, and any finite temperature, there is at least one
distribution over the energy spectrum of that Hamiltonian, that
corresponds to that temperature.

In other words, under the assumption of universality, given a
collection, $\{\rho_1, \ldots, \rho_n\}$, of distributions at the
same temperature, and any Hamiltonian, $H_{n+1}$, of a new system,
then it must always be possible to find a distribution $\rho_{n+1}$
over the new system, such that $\rho_i \otimes \rho_{n+1}$ is
passive, for all $1 \leq i \leq n$.
\end{enumerate}

We can use these properties to establish that the ratio of the
probabilities of any two energy levels must be a single valued
function of their energy difference, and the temperature.

First suppose we have a given system, with two energy levels
separated by the gap $E_1-E_0=\Delta$, and the ratio of whose
probabilities is given by $p_1/p_0=\Pi$.  If we consider two levels,
$i$ and $j$, of any other system, then comparison of the $\ket{E_1
E_i}$ and $\ket{E_0 E_j}$ levels yields:
\begin{equation}
E_i - E_j \geq \Delta \Leftrightarrow \frac{p_i}{p_j} \leq \Pi
\end{equation}

Now we consider all Hamiltonians that contain levels with the energy
gap $\Delta$.  This will yield maximum and minimum values of $\Pi$
for that energy gap, and so:
\begin{eqnarray}
E_i - E_j \geq \Delta & \Leftrightarrow & \frac{p_i}{p_j} \leq \Pi_{min} \nonumber \\
E_i - E_j \leq \Delta & \Leftrightarrow & \frac{p_i}{p_j} \geq
\Pi_{max}
\end{eqnarray}
As we can do this for all values of $\Delta$, we generate two
functions $\Pi_{max}(\Delta)$ and $\Pi_{min}(\Delta)$.  A little
thought shows, that given for all $\Delta^\prime \geq \Delta$, then
$\Pi_{max}(\Delta^\prime) \leq \Pi_{min}(\Delta)$, these functions
must both be piecewise continuous.

We next demonstrate that $\Pi_{max}(\Delta) = \Pi_{min}(\Delta)$.
Consider a Hamiltonian with four non-degenerate energy levels, for
which the energy gaps between the lowest and the three higher are
$\Delta-\delta_1$, $\Delta-\delta_2$ and $\Delta+\delta_3$,
respectively, such that $\delta_1 > \delta_2$ and $\delta_1,
\delta_2, \delta_3 >0$.  The inverse ratio of the probability of the
lowest of the energy level, to the $\Delta-\delta_1$ energy level,
is $\Pi_1$, to the $\Delta-\delta_2$ energy level, is $\Pi_2$, and
to the $\Delta+\delta_3$ energy level, is $\Pi_3$.

Now consider the product of two systems with the four energy levels.
Comparing the $\ket{E_2,E_2}$ with the $\ket{E_1,E_3}$ levels, if
$\delta_3 - \delta_1 \geq 2 \delta_2$ then $(\Pi_2)^2 \leq \Pi_1
\Pi_3$.

By the definition of $\Pi_{max}(\Delta)$ and $\Pi_{min}(\Delta)$, as
we vary through Hamiltonians so that $\delta_1, \delta_2, \delta_3
\rightarrow 0$, we have
\begin{eqnarray}
\Pi_1, \Pi_2 &\rightarrow & \Pi_{max}(\Delta) \nonumber \\
\Pi_3 &\rightarrow & \Pi_{min}(\Delta)
\end{eqnarray}
Provided we maintain $\delta_3 - \delta_1 \geq 2 \delta_2$, such as
by $\delta_2 = \frac{1}{4}\left(\delta_3 - \delta_1\right)$, then
\begin{equation}
\Pi_{max}(\Delta)^2 \leq \Pi_{max}(\Delta) \Pi_{min}(\Delta)
\end{equation}
which gives $\Pi_{max}(\Delta) \leq \Pi_{min}(\Delta)$, but
$\Pi_{max}(\Delta) \geq \Pi_{min}(\Delta)$, by definition, so
$\Pi_{max}(\Delta) = \Pi_{min}(\Delta)=\Pi(\Delta)$.

We now have demonstrated that the property of universality of
temperature, requires that there exists, for each value $T$, of
temperature, a function $\Pi_T(\Delta)$, such that
\begin{equation}
\frac{p_i}{p_j}=\Pi_T(E_i-E_j)
\end{equation}
It is a simple matter to deduce the unique function that satisfies
this. Considering a third level
\begin{equation}
\frac{p_k}{p_j}=\Pi_T(E_k-E_j)=\frac{\Pi_T(E_i-E_j)}{\Pi_T(E_i-E_k)}
\end{equation}
Writing $\Delta_1=E_k-E_j$ and $\Delta_2=E_i-E_k$, gives
\begin{equation}
\Pi_T(\Delta_1+\Delta_2)=\Pi_T(\Delta_1)\Pi_T(\Delta_2)
\end{equation}
We find a variation on the Darboux relationship\cite{Darboux1880},
which has the solution:
\begin{equation}
\frac{p_i}{p_j}=e^{-\beta(T)(E_i-E_j)}
\end{equation}
We can rearrange this
\begin{equation}
p_i e^{\beta(T)E_i}=p_j e^{\beta(T)E_j}=\lambda
\end{equation}
where $\lambda$ is a constant, which by normalisation of the
probabilities, is
\begin{equation}
\lambda=\frac{1}{\sum_i e^{-\beta(T)E_i}}
\end{equation}

Transitivity therefore yields the canonical distribution, but only
if we supplement it with the requirement that temperature be
universal.

\end{enumerate}

\subsection{Comments}
The concept of passivity is clearly closely related to the concept
of equilibrium. However, we have not made any assumptions regarding
whether systems, not in equilibrium, must evolve into a state of
equilibrium, or whether systems, in equilibrium, may spontaneously
be found out of equilibrium.  The temporal asymmetry associated with
the concept of equilibrium\cite{BU2001}, which is so problematical
for the relationship of statistical mechanics to thermal physics,
has not been assumed.

It appears we do not need to assume that isolated systems tend to
equilibrium to understand what a \textit{thermal} state must be. The
only sense of equilibrium that we may be argued to have used here,
is a sense in which two states at the same temperature may be
regarded as being in equilibrium with each other.  This is a
relationship between systems, not property of individual systems.

Having said that, if the result of thermalising a system did not
lead to a passive distribution, then clearly we could create
perpetual motion: simply isolate the thermalised system, extract the
available energy, then restore thermal contact and allow the system
to return to the thermalised state.

Another sense of equilibrium, that a distribution is in equilibrium
if it is constant in time, also leads to the conclusion that the
distribution must be diagonalised in the energy eigenbasis.  Of
course, we have seen that the concept of passivity is also
sufficient to deduce the density matrix diagonalises the energy
eigenbasis.

That the compositional property rules against the microcanonical
distribution was noted, in passing, by Gibbs\cite{Gib1902}[pg.
170-4].  It is key (although not obviously so) to Szilard's
derivation of the canonical distribution for classical
systems\cite{Szi1925}.

It is the factorisability of the joint probability distribution of
the combined state that leads uniquely to the canonical
distribution. This is deeply related to the development of the
statistical mechanical account of thermal phenomena.  Non-extensive
entropies must, at least implicitly, deny this property.  Both the
extensivity of Gibbs entropy and the additivity of Shannon
information are, formally, closely related to this.  The extensivity
of entropy is a somewhat abstract concept, which has something of
the status of convention even in classical phenomenological
thermodynamics. Shannon information has no obvious a priori
relationship to thermal phenomena and has played no role in this
derivation.  The compositional property of temperature is a
verifiable physical property of thermal states.

\subsection{The Ideal Gas Scale}
The temperature \textit{scale} is largely a matter of convention.
Consistency amongst different operationally defined temperatures
requires that any two thermal states considered to be at the same
temperature with respect to one scale, are at the same temperature
with respect to any other scale.  It is usually also assumed that
the primitive ordering of \textit{hotter than/colder than} is also
preserved across all temperature scales.  This would mean that any
temperature scale can be expressed as a monotonic function of any
other scale.

We could choose to use $\beta(T)$ as the definition of our
statistical temperature scale.  Provided $\beta$ is a single valued
function, it satisfies the consistency requirement.  To verify the
ordering relationship, we will identify a particular operationally
defined temperature scale and find how $\beta(T)$ varies with that.

We will make the choice of the ideal gas scale.  It is empirically
observed that, for a number of gases, the relationship:
\begin{equation}
PV=NRT
\end{equation}
appears to hold, where the number of moles of the gas is $N$ and $R$
is the molar gas constant.  It is hypothesised this holds exactly
for \textit{ideal} gases.

Standard textbook analysis of a canonically distributed system,
gives us the result that an ideal $n$ quantum gas confined to a box
of volume $V$, exerts a mean pressure $P$ on the walls of the box
of:
\begin{equation}
PV\beta(T)=n
\end{equation}
The Boltzmann constant $k=RN/n$ gives
\begin{equation}
\beta(T)=\frac{1}{kT}
\end{equation}
As this is a single valued, monotonic function of $T$, it follows
$\beta(T)$ is a good temperature scale, provided the ideal gas
scale, $T$, is a good temperature scale.

\section{Statistical Thermal States}

We have identified thermal states as uniquely represented in
statistical mechanics by canonical probability distributions.  The
canonical probability distribution is characterised by a single
parameter, $\beta(T)$, which can be related to the reciprocal of the
ideal gas scale.  We will now consider what we can deduce, solely
from the identification of thermal states as canonical probability
distributions, using the techniques of statistical mechanics.

\subsection{Mean flow
of energy} We can state immediately the consequence of Equation
\ref{eq:temptheorem}:
\begin{equation}
\Delta H_1 \left(\beta_2 - \beta_1\right)\geq 0
\end{equation}
If two thermal states interact, with \textit{ different} parameters
$\beta_1$ and $\beta_2$, then if $\beta_1 > \beta_2$, the mean flow
of energy, $\Delta H_1  \leq0$, can only
 be from system 2 to system 1.

\subsection{Thermal cycles}
Next consider a system, with an arbitrary probability distribution,
$\rho^{(0)}$, initially uncorrelated or interacting with any other
system.  The system is now brought into successive contact with a
series of systems in thermal states, where the state of system $i$
is parameterised by $\beta_i$.  After each contact has ceased, the
expectation value for the interaction energy with the $i^{th}$
system is zero.  The internal Hamiltonians for the canonical states
systems are constant in time.

Let $\rho^{(i)}$ be the marginal probability distribution of the
system after interacting with the $i^{th}$ thermal system.  Let
$\rho_i(\beta_i)$ be the initial probability distribution of the
$i^{th}$ thermal system, and  $\rho_i^\prime$ be the marginal
probability distribution afterwards.  Let $\Delta H_i$ represent the
mean energy flow into the $i^{th}$ thermal system.  We do not assume
any of the systems are canonically distributed after the
interaction.

Now, purely from the mathematical properties of canonical
distributions (Section \ref{s:distributions}) we can state:
\begin{eqnarray}
G\func{\rho^{(i-1)}}+G\func{\rho_i(\beta_i)} &\geq&
G\func{\rho^{(i)}}+G\func{\rho_i^\prime} \\
G\func{\rho_i^\prime}-G\func{\rho_i(\beta_i)} & \geq & -\beta_i
\Delta H_i
\end{eqnarray}

Adding these together and summing over all interactions we get
\begin{equation}
G\func{\rho^{(0)}}-G\func{\rho^{(f)}} \geq - \sum_i \beta_i \Delta
H_i
\end{equation}
from which it follows that, if the series of interactions is such
that it returns the systems final marginal probability distribution
to its initial probability distribution $\rho^{(f)}=\rho^{(0)}$
then:
\begin{equation}\label{eq:betaclaus}
\sum_i \beta_i \Delta H_i \geq 0
\end{equation}
No physical interpretation need be placed upon the Gibbs-von Neumann
measure $G$, to derive these results.  We are deducing a property
(Equation \ref{eq:betaclaus}) of canonical distributions, under
Hamiltonian evolution.

We have not needed to assume that the system Hamiltonian returns to
it's initial value.  If this is the case, the mean work performed
over the course of the cycle is
\begin{equation}
\Delta W=\sum_i \Delta H_i
\end{equation}

\subsection{Statistical thermalisation}
Now consider a system with an arbitrary probability distribution,
$\rho^{(0)}$, initially uncorrelated or interacting with any other
system, and a Hamiltonian $H$.  The system is now brought into
successive contact with a series of systems in thermal states, where
each system has the same parameter $\beta$.  Let $\rho^{(i)}$ be the
marginal probability distribution of the system after interacting
with the $i^{th}$ thermal system, and let $\rho(\beta)$ be the
thermal state for the system with parameter $\beta$.

\begin{equation}
G\func{\rho(\beta)}+\beta \mean{H}_{\rho(\beta)} \leq
G\func{\rho^{(i+1)}}+\beta \mean{H}_{\rho^{(i+1)}} \leq
G\func{\rho^{(i)}}+\beta \mean{H}_{\rho^{(i)}} \leq
G\func{\rho^{(0)}}+\beta \mean{H}_{\rho^{(0)}}
\end{equation}

As each equality holds only if the system is canonically distributed
$\rho(\beta)$ we conclude that, if there is no physical cause that
prevents it, a system with any arbitrary probability distribution
can be brought arbitrarily close to a canonical probability
distribution, with the parameter $\beta$, by a sufficiently large
number of such contacts. The system becomes thermalised.

Possible physical causes that prevent complete thermalisation
include:
\begin{enumerate}
\item Transitions are not permitted between different regions of the
state space.  Let the different regions of state space be
represented by the complete set of non-overlapping projectors $K_i$
onto those regions, the partial thermalisation of the density matrix
$\rho$, will be:
\begin{equation}
\rho^\prime(\beta)=\sum_i \trace{K_i \rho K_i}\frac{e^{-\beta K_i H
K_i }}{\trace{e^{-\beta K_i H K_i }}}
\end{equation}
\item Only part of the region of the state space represents the
system being in thermal contact (i.e. the interaction Hamiltonian is
zero for some portion of the system state space) and there are no
transitions out of that region. If $K_\alpha$ projects onto the
isolated region, and $K_\beta$ onto the thermal contact region,
partial thermalisation will lead to:
\begin{equation}
\rho^\prime(\beta)=K_\alpha \rho K_\alpha + \trace{K_\beta \rho
K_\beta}\frac{e^{-\beta K_\beta H K_\beta }}{\trace{e^{-\beta
K_\beta H K_\beta }}}
\end{equation}
\end{enumerate}

At the risk of getting repetitive, we restate: no physical
interpretation is placed upon the Gibbs measure $G$ to derive these
results.  It is used simply to establish general mathematical
properties of the evolution of distributions under Hamiltonian
evolutions and of interactions with thermal states.  The properties
of these interactions can be understood without needing to
physically interpret the Gibbs measure.  They are properties of
Hamiltonian evolutions.

\subsection{Heat Baths}
It will now be convenient to identify heat baths.  A heat bath is
simply a large system, with many degrees of freedom, in a thermal
state.  With many degrees of freedom it can be treated as having a
large number of subsystems.  No work is ever performed upon a heat
bath.

When a system interacts with a heat bath, it generally interacts
only with one of the subsystems. What happens following that depends
upon the details of the heat bath, the interactions between the
subsystems and whether continued interaction with the heat bath
involves continued interaction with the same, or a different,
subsystem.  We will use the symbol $\Delta Q$ to refer to mean
energy flows into a heath bath, and will refer to these as mean heat
exchanges.

\subsubsection{Ideal Heat baths}
An `ideal' heat bath, which is one for which the subsystems are
non-interacting, and an individual subsystem is never encountered
twice.  All the subsystems are canonically distributed with the same
$\beta$ parameter, and it is assumed there are no internal
microscopic correlations.  This means a system brought into contact
with an ideal heat bath experiences a succession of contacts with
independant canonically distributed systems at the same $\beta$
parameter.  It will become thermalised. An ideal heat bath may be be
treated as a Large Uncorrelated Canonical Assembly (Section
\ref{ss:luca}) which is identical to a Large Completely Passive
Assembly. The interaction with each subsystem will be for a very
short time\footnote{Care needs to be taken regard a limiting case of
infinitesimally short interactions or the quantum Zeno effect will
prevent thermalisation at all.}. From the completely passive
properties of thermal systems, we can state immediately it is
impossible to extract work from any number of ideal heat baths at
the same temperature.

\subsubsection{Real Heat baths}
Of course, as soon as the systems have interacted, correlations
develop and do not disappear.  Subsystems of real heat baths
interact with each other.  Subsystems may be re-encountered.

To judge the consequences of these requires real models of physical
heat baths. For example, one property of weakly interacting
subsystems\cite{Par89b} is a tendency for correlations to become
``spread out'', so that the correlation between the system and the
heat bath subsystem is reduced by the weak interaction amongst the
heat bath subsystems.

Real heat baths do not behave exactly as ideal heat baths. The
properties we are going to derive based upon ideal heat baths will
not, therefore, be strictly applicable to interactions with real
heat baths.  The extent to which the behaviour of real heat baths
differs from that of ideal heat baths can only be decided by
examining physical models of the real heat baths.

That real heat baths do not behave exactly as ideal heat baths is
not a fundamental problem for statistical mechanics.  Statistical
mechanics should not be required to prove real heat baths behave as
ideal heat baths - after all, they don't!  When considering real
heat baths, with real physical interactions, statistical mechanics
is required to accurately describe the actual behaviour of those
real heat baths, including how they \textit{deviate} from being
ideal heat baths, given the appropriate description of the real
physical interaction.

\subsection{Statistically isothermal operations}
We now consider the limiting cases of interactions with an ideal
heat bath.  When a thermal system, with internal Hamiltonian $H$ is
brought into contact with an ideal heat bath with a fixed $\beta$
parameter, the system is kept in a thermalised
state\footnote{Provided the Hamiltonian is varied only slowly.}. The
resulting canonical distribution gives a density matrix which is
always diagonalised with respect to the system Hamiltonian.  As the
system Hamiltonian is varied, the mean work performed is
\begin{eqnarray}
\Delta \mean{W}&=&\int_0^\tau \trace{\partime{H}\frac{e^{-\beta
H}}{\trace{e^{-\beta H}}}}dt\\
 &=&\int_0^\tau \sum_n \partime{E_n} \frac{e^{-\beta
E_n}}{\sum_n e^{-\beta E_n}} dt
\end{eqnarray}
and the mean heat exchanged is
\begin{equation}
\Delta \mean{Q}=\int_0^\tau \sum_n E_n \partime{
}\left(\frac{e^{-\beta E_n}}{\sum_n e^{-\beta E_n}}\right) dt
\end{equation}
The work may be re-expressed as
\begin{equation}
\Delta \mean{W}=-\frac{1}{\beta}\ln\func{\frac{Z(\tau)}{Z(0)}}
\end{equation}
where
\begin{equation}
Z(t)=\trace{e^{-\beta H(t)}}
\end{equation}

\section{Statistical Entropy}
The results of the previous section have not, at any point, depended
upon the identification of  $-G$ with thermodynamic entropy.  The
Gibbs-von Neumann measure of a distribution has been treated simply
as a convenient calculation tool, and has not been attributed any
physical significance.  The ``entropy-like'' qualities of $G$ that
have been used are simply mathematical properties of any
distribution function under Hamiltonian evolutions.  No physical
interpretation has been placed upon them, nor was needed to use
them.

Nevertheless we have managed to derive the inequality
\begin{equation}
\sum_i \frac{\Delta Q_i}{T_i} \geq 0
\end{equation}
where $Q_i$ is the mean heat flow, into a thermal system at
temperature $T_i$, over a closed cycle.  The mean work performed
over the course of the cycle is
\begin{equation}
\Delta W=\sum_i \Delta Q_i
\end{equation}

Consider a few special cases of this, for heat baths:
\begin{enumerate}
\item If there is a single heat bath:
\begin{equation}
\Delta Q \geq 0
\end{equation}
the mean flow of energy \textit{ must} be into the heat bath.
\item If there are two heat baths, and the mean work requirement for the process is zero, $\Delta
Q_1=-\Delta Q_2=\Delta Q$:
\begin{equation}
\Delta Q\left(\frac{1}{T_1}-\frac{1}{T_2}\right) \geq 0
\end{equation}
$\Delta Q > 0$ if and only if $T_1 \leq T_2$.  The mean flow of heat
into the first heat bath can be positive only if the first heat bath
is colder than the second.
\item If there are two heat baths,
\begin{equation}
\frac{\Delta W}{\Delta Q_1} \leq 1 - \frac{T_2}{T_1}
\end{equation}
Which (with due regard for changes in sign to both $\Delta W$ and
$\Delta Q_1$) shows the maximum efficiency, in terms of mean work
extracted over mean heat extracted, for a heat engine.
\end{enumerate}

\subsection{Thermodynamic entropy}
We will now, for the first time, consider phenomenological
thermodynamics.  The primitive exposition of the concept of
thermodynamic entropy here clearly lacks the careful rigour of such
works as \cite{GB1991,LY1999} and indeed differs greatly from them.
It is closer to such textbook expositions such
as\cite{Fer1937,Adk68}.  Our reasons for this are simple: it is not
obvious what the statistical mechanical generalisation of
thermodynamic entropy should be and there is no universal agreement
on what properties of thermodynamic entropy are the ones to select
in developing this generalisation (or even if such a generalisation
is necessary).  Our approach will be to focus upon the arguments
that typically motivate supposing that there is such a thing as
thermodynamic entropy in the first place and see how these arguments
apply to statistical mechanics.

The three cases in the previous section may be compared to three
versions of the Second Law of Thermodynamics
\begin{enumerate}
\item No process is possible whose sole result is the extraction of
heat from a heat bath and its conversion into work.
\item No process is possible whose sole result is the transfer of
heat from a colder to a hotter heat bath.
\item No process is possible whose sole result is the extraction of
heat $Q_1$ from a heat bath at temperature $T_1$ and the deposit of
heat $Q_2$ in a heat bath at temperature $T_2<T_1$, extracting the
remainder $W=Q_1-Q_2$ with efficiency $\eta=W/Q_1$ greater than
$1-T_2/T_1$.
\end{enumerate}

An equivalent expression of the Second Law would be:
\begin{enumerate} \setcounter{enumi}{3}
\item No process is possible, whose sole result is the use of work to
transfer heat between heat baths, such that the heat deposited in
the $n^{th}$ heat bath, at temperature $T_n$, is $Q_n$ and
\begin{equation} \label{eq:clausiusII}
\sum_n \frac{Q_n}{T_n} < 0
\end{equation}
\end{enumerate}

Now, the final expression leads to the following result:
\begin{quote}
If there exists a process, which takes a system from state A, to a
system at state B, depositing heats $Q^{(1)}_m$ in heat baths at
temperatures $T^{(1)}_m$, then there can be no process, which takes
a system in state B, to a system in state A, depositing heat
$Q^{(2)}_n$ in heat baths at temperatures $T^{(2)}_n$, unless
\begin{equation}\label{eq:bothways}
\sum_m \frac{Q^{(1)}_m}{T^{(1)}_m}+\sum_n
\frac{Q^{(2)}_n}{T^{(2)}_n} \geq 0
\end{equation}
\end{quote}

As this result must hold for \textit{ any} two states A and B in any
combination of processes, this is equivalent to the statement:
\begin{quote}
There exists a single valued property, $S$, of a state such that, if
there exists a process, which takes a system from state A, to a
system at state B, depositing heats $Q^{(1)}_m$ in heat baths at
temperatures $T^{(1)}_m$, then
\begin{equation}\label{eq:defthentropy}
S\func{B}-S\func{A} \geq - \sum_m \frac{Q^{(1)}_m}{T^{(1)}_m}
\end{equation}
\end{quote}
The value of this property $S$ is not yet uniquely defined.  There
may be many functions which satisfy this requirement.

If there also exists a process, which takes a system from state B,
to a system at state A, depositing heats $Q^{(1)}_n$ in heat baths
at temperatures $T^{(1)}_n$, such that
\begin{equation}
\sum_m \frac{Q^{(1)}_m}{T^{(1)}_m}+\sum_n
\frac{Q^{(2)}_n}{T^{(2)}_n} = 0
\end{equation}
then $S$ is uniquely defined\footnote{Up to an additive constant, of
course!} by:
\begin{equation}
S\func{B}-S\func{A} = -\sum_m \frac{Q^{(1)}_m}{T^{(1)}_m}
\end{equation}
Identifying the value of $S$ for all states then requires us to find
processes in both directions which are able to complete the cycle
with the requisite minimal transfer of heat to heat baths.

If we can identify the value of $S$ for all states then it follows
\begin{quote}
There is no process which takes a system from state A, to a system
at state B, depositing heats $Q^{(1)}_m$ in heat baths at
temperatures $T^{(1)}_m$, for which
\begin{equation}
S\func{B}+ \sum_m \frac{Q^{(1)}_m}{T^{(1)}_m} < S\func{A}
\end{equation}
\end{quote}
and finally, if we identify the change in the value of $S$ for a
heat bath with $Q/T$, we get
\begin{quote}
There is no process for which
\begin{equation}
\sum \Delta S < 0
\end{equation}
\end{quote}

The quantity $S$ is the thermodynamic entropy of the state.

\subsection{Statistical mechanical generalisation of thermodynamic entropy}
The first thing we may wonder is whether there is any need to
introduce the concept of a statistical mechanical entropy. The
existence of the non-decreasing function of state is equivalent to
the various operational statements of the second law, about the
absence of certain kinds of processes.  One of these statements must
be introduced into the axiomatic structure of thermodynamics to be
able to deduce results as, whichever statement is chosen as the
appropriate second law axiom, otherwise it cannot be deduced.

For statistical mechanics, we \textit{can} deduce the statistical
equations from the properties of Hamiltonian dynamics, probability
calculus and the existence of thermal states.  There is no need to
introduce a new axiom.  Nevertheless, the tremendous utility of the
thermodynamic entropy function in developing phenomenological
thermodynamics should suggest to us that such a function may be
useful.  Perhaps it may be possible, in principle, to develop
phenomenological thermodynamics without introducing an entropy
function, but instead, for example, rely solely upon the Kelvin
formulation of the second law, but it would seem needlessly
difficult to do so.

If we decide to introduce such a function, let us consider the
phenomenological laws which motivate introducing it.  The first
thing to note is that they are all \textit{false.}  It is possible
to have processes whose sole result is to convert heat to work (just
with probability less than one).  It is possible to have processes
whose sole result is to transfer energy from a colder to a hotter
heat bath (just with probability less than one).

This has a profound consequence for the development of a statistical
mechanical entropy.  The justification for introducing entropy as a
single valued property of state comes from Equation
\ref{eq:bothways} above. This justification \textit{ does not hold}
for microstates.

Hamiltonians exist which can transform any microstate into any other
microstate, while extracting arbitrarily large amounts of heat from
a heat bath and converting it into work, so long as we are prepared
to accept an arbitrarily low probability of the process occurring.
No single valued entropy function could be deduced from this
attempt.

Attempting to fix this by demanding that the process can occur with
certainty, we find that there is always some Hamiltonian evolution
on the state space, which can perform a transformation between any
two given microstates, without any exchange of energy with a heat
bath. The same conclusion is reached even if we demand only that the
\textit{mean} transfer of heat to the heat baths be zero. The
entropy difference between any two microstates is zero.

\subsubsection{Defining the statistical mechanical entropy}
The steps that might lead us to try to deduce the existence of an
entropy function as a function of the microstate of a system are
flawed. However, there was no reason to take those steps.  The
search for new axioms to introduce, to represent a statistical
mechanical second law, is unnecessary.  Statistical mechanics has
already enabled us to deduce the property:
\begin{quote}
No process is possible, starting with any probability distribution
over a system, whose sole result is to return the system to it's a
marginal probability distribution equal to it's initial
distribution, and to transfer mean quantities of energy $\Delta Q_i$
into systems initially uncorrelated and canonically distributed with
parameters $\beta_i$, where
\begin{equation}
\sum_i \beta_i \Delta Q_i < 0
\end{equation}
\end{quote}
Note that this deduction, as expressed, is a direct consequence of
Hamiltonian dynamics and does not depend upon any identification of
thermal states, temperatures or heat baths.

Accepting the identification of thermal states as canonical
distributions, and the state of a system being here defined as a
Hilbert space, Hamiltonian operator on that space and probability
distribution over the space, it follows:
\begin{quote}
There exists a single valued property, $S$, of the states of systems
such that, if there exists a process, which takes a system from
state A, to a system at state B, depositing mean heats $Q^{(1)}_m$
in thermal systems at temperatures $T^{(1)}_m$, then
\begin{equation}
S\func{B}-S\func{A} \geq - \sum_m \frac{Q^{(1)}_m}{T^{(1)}_m}
\end{equation}
\end{quote}
We define this property as the statistical mechanical generalisation
of thermodynamic entropy.  We will now calculate its value.
\subsubsection{Deriving statistical mechanical entropy}
In order to fix the entropy difference between two states, it is
necessary to find processes between the states in both directions,
where:
\begin{equation}
\sum_m \frac{Q^{(1)}_m}{T^{(1)}_m}+\sum_n
\frac{Q^{(2)}_n}{T^{(2)}_n} = 0
\end{equation}
For phenomenological thermodynamics, this involves reversible,
quasistatic processes.  These processes do not actually exist in
reality.  They are limiting processes - they are not necessarily
attainable, but there is no physical reason one cannot get
arbitrarily close to them.  However, they are generally considered
to only be possible for systems in thermal equilibrium.

In statistical mechanics we have a more general notion of process
available: Hamiltonian dynamics.  From this we can construct
reversible processes\footnote{Please note, the specific Hamiltonians
used are provided solely to demonstrate the existence of concrete
examples.}, as a limiting case.
\begin{enumerate}
\item {Isothermal processes}

We first identify the entropy change for an isothermal process.
Taking an ideal heat bath as the limit of a real heat bath, we
assume we can get arbitrarily close to an ideal heat bath process.
An isothermal process in contact with a single heat bath at
temperature $T$, requires mean work equal to:
\begin{equation}
\Delta \mean{W}=-kT\ln\func{\frac{Z(\tau)}{Z(0)}}
\end{equation}
where
\begin{equation}
Z(t)=\trace{e^{-H(t)/kT}}
\end{equation}
The mean change in energy is
\begin{equation}
\Delta E=\trace{H(\tau)}-\trace{H(0)}
\end{equation}
The mean heat transferred to the heat bath is
\begin{equation}
\Delta Q=\Delta W - \Delta E
\end{equation}
Substituting $E_n=-kT \ln Z - kT \ln \func{p_n}$ for canonical
thermal systems gives
\begin{equation}
\Delta Q = kT\left(\sum_n p_n(\tau)\ln\func{p_n(\tau)}-\sum_n
p_n(0)\ln\func{p_n(0)}\right)
\end{equation}

\item{Non-isothermal processes}

While the above relationship may be well known for quasistatic,
isothermal processes, we must consider a more general process.
Suppose we have a system that starts in an arbitrary, uncorrelated
state
\begin{equation}
\rho=\sum_n p_n \proj{\alpha_n}
\end{equation}
with Hamiltonian $H_i$.  We wish to find the entropy difference with
another state
\begin{equation}
\rho^\prime=\sum_n p^{\prime}_n \proj{\beta_n}
\end{equation}
with Hamiltonian $H_f$.  We need to find a reversible Hamiltonian
process to achieve this.

 We break the evolution into three stages, and give the Hamiltonian
 for each stage.

\begin{enumerate}
\item{$0<t<\tau_1$}

Keeping the system isolated

\begin{eqnarray}
H_{A}&=&H_i \left[\cos^2\left(\frac{\pi t}{2 \tau_1}\right)-
\sin^2\left(\frac{\pi t}{\tau_1}\right)\right] + H_1
\left[\sin^2\left(\frac{\pi t}{2 \tau_1}\right)-
\sin^2\left(\frac{\pi t}{\tau_1}\right)\right] \eqnwrap -\frac{2
\imath \hbar}{\tau_1} \sin^2\left(\frac{\pi t}{\tau_1}\right) \ln
\func{\sum_n \kb{\gamma_n}{\alpha_n}}
\end{eqnarray}
with
\begin{equation}
H_1=\sum_n -kT \ln(p_n) \proj{\gamma_n}
\end{equation}
The effect of this evolution is to leave the system in the canonical
state
\begin{equation}
\rho=\frac{e^{-H_1/kT}}{\trace{e^{-H_1/kT}}}=\sum_n p_n
\proj{\gamma_n}
\end{equation}

\item{$\tau_1<t<\tau_2$}

Bring the system into thermal contact with a heat bath at
temperature $T$, and isothermally, quasi-statically change the
Hamiltonian to
\begin{equation}
H_2=\sum_n -kT \ln(p^{\prime}_n) \proj{\gamma_n}
\end{equation}
At the end of this process, the system is in the state
\begin{equation}
\rho=\frac{e^{-H_2/kT}}{\trace{e^{-H_2/kT}}}=\sum_n p^{\prime}_n
\proj{\gamma_n}
\end{equation}

\item{$\tau_2<t<\tau_f$}

Isolate the system again and the final Hamiltonian is
\begin{eqnarray}
H_{B}&=&H_2 \left[\cos^2\left(\frac{\pi t}{2
(\tau_f-\tau_2)}\right)- \sin^2\left(\frac{\pi
t}{\tau_f-\tau_2}\right)\right] \\ && + H_f
\left[\sin^2\left(\frac{\pi t}{2 (\tau_f-\tau_2)}\right)-
\sin^2\left(\frac{\pi t}{\tau_f-\tau_2}\right)\right] -\frac{2
\imath \hbar}{\tau_f-\tau_2} \sin^2\left(\frac{\pi
t}{\tau_f-\tau_2}\right) \ln \func{\sum_n \kb{\beta_n}{\gamma_n}}
\nonumber
\end{eqnarray}
which produces the desired final matrix $\rho^\prime$.
\end{enumerate}

Stages (a) and (c) are isolated, so involve no heat exchange.
Changes in internal energy are entirely through work performed upon
the system. In the limit of ideal heat baths and slow isothermal
processes the net heat exchange is
\begin{equation}
\Delta Q = kT\left(\sum_n p^\prime_n\ln\func{p^\prime_n}-\sum_n
p_n\ln\func{p_n}\right)
\end{equation}

\end{enumerate}

The limiting process can clearly take place in either direction.  We
can therefore deduce that, for any two density matrices $\rho$ and
$\rho^\prime$, the difference in the statistical mechanical
generalisation of their \textit{ thermodynamic} entropy is:

\begin{equation}
S\func{\rho^\prime}-S\func{\rho} = - k\left(\sum_n
p^\prime_n\ln\func{p^\prime_n}-\sum_n p_n\ln\func{p_n}\right)
\end{equation}

As this relationship
\begin{equation}
S\func{\rho^\prime}+ k\sum_n p^\prime_n\ln\func{p^\prime_n}=
S\func{\rho}+k\sum_n p_n\ln\func{p_n}
\end{equation}
must hold for any density matrices, and with any eigenstates, then

\begin{equation}
S\func{\rho}=-k\trace{\rho \ln \func{\rho}}+c
\end{equation}

where $c$ is a universal additive constant which can be set to zero
by convention. The Gibbs-von Neumann entropy is deduced to be the
correct generalisation of the thermodynamic entropy, for statistical
mechanics.

\subsection{Optimal processes}
The physical principle expressed by the Gibbs-von Neumann entropy
is:
\begin{quote}
There is no process which takes a system from state A, to a system
at state B, depositing mean heats $Q_m$ into thermal systems at
temperatures $T_m$, for which
\begin{equation}
S\func{B}-S\func{A} + \sum_m \frac{Q_m}{T_m} < 0
\end{equation}
\end{quote}

The limiting process, where
\begin{equation}
S\func{A} =S\func{B}+ \sum_m \frac{Q_m}{T_m}
\end{equation}
is the \textit{ optimal} process.  It is the process, which, on
average, generates the least heat.

The thermal systems into which heat is transferred are not
necessarily ideal heat baths.  All that is assumed is that they are
initially uncorrelated to other systems and that they are described
by a canonical probability distribution.

There is no special reason why the \textit{ mean} generation of heat
is important.  Other criteria may be considered. In some physical
circumstances other properties might be more important.  One might
wish to find the optimal process according to some other criteria,
such as a minimax criteria (minimising the maximal cost).
Statistical mechanics provides the tools for doing this.

If we search for a process which minimises the mean generation of
heat in thermal systems, we are lead to the statistical mechanical
entropy as the quantity which characterises the optimal process.  If
we search for a process by some other criteria, we will find
different quantities of interest and different processes.  These
would, necessarily, involve at least as much heat generation, on
average, but would outperform the optimal entropic process according
to some other criteria.
\subsection{Consistency}
Can we be sure this definition of thermodynamic entropy is
consistent?  Once the identification of statistical temperature and
the gas scale has been made, it is possible to derive the entropy
relationships directly from the property of the Gibbs-von Neumann
measure $G$, and the properties of canonical distributions.  It
holds for all processes, because of two properties of Hamiltonian
evolutions:

\begin{enumerate}
\item For any process, which starts with uncorrelated
distributions over a number of systems, the change in the value of
$G$ for each marginal distribution gives

\begin{equation}
\sum_i \Delta G_i \leq 0
\end{equation}

\item For any system, initially canonically distributed
\begin{equation}
\Delta G + \beta \Delta \mean{H} \geq 0
\end{equation}
\end{enumerate}

We did not go directly from these properties to the identification
of $-k G$ with the thermodynamic entropy as we wished to justify
precisely what purpose thermodynamic entropy is intended to fulfil.
Having done so, and demonstrated that $-k G$ is the correct value,
we can now see that the result
\begin{equation}
\sum_i \Delta S_i \geq 0
\end{equation}
must hold, provided that the evolution is Hamiltonian and that
initially independant systems are uncorrelated.

\subsection{Non-equilibrium statistical mechanics}

The entropy has been derived for arbitrary probability
distributions, not only for systems in thermal equilibrium.
Phenomenological thermodynamic entropy is frequently regarded as
only being well defined for states in thermal equilibrium.  How is
it that statistical mechanics could do better?

There is a subtlety involved in the temperature in Equation
\ref{eq:clausiusII}.  This gives the temperatures of the thermal
systems, into which heat is transferred.  It is not, directly, the
temperature of the system from which heat is being expelled.  It is
the temperature of the heat baths, not the temperature of the system
undergoing the cyclic process.  The phenomenological thermodynamic
entropy function, defined by Equation \ref{eq:defthentropy}, is
therefore defined for \textit{ all} states, whether they are in
thermal equilibrium or not.

To identify the actual value of the entropy difference between two
states, it is necessary to identify reversible processes in both
directions.  For phenomenological thermodynamics, such process are
only known in the limiting case of quasistatic processes on states
in thermal equilibrium.  Although the temperature $T$ that appears
in the summation is strictly the temperature of the \textit{ heat
bath}, the change in $S$ can usually only be uniquely identified
when the system is kept in thermal equilibrium at the same $T$ as
the heat bath (see \cite{Fer1937}[Chapter IV, Section 11]).  This
can lead to the claim that thermodynamic entropy is only
well-defined for systems in thermal equilibrium.

While this may, arguably, be true for phenomenological
thermodynamics, there is no reason to insist that it should also be
true for statistical mechanics.  Statistical mechanics comes with a
well defined notion of processes - Hamiltonian evolution - even when
systems are not in equilibrium.  In statistical mechanics, there is
no need to artificially restrict the domain of validity of the
entropy function.

\section{Conclusions}
We have come to the conclusion that the Gibbs-von Neumann entropy is
the appropriate statistical mechanical generalisation for
thermodynamic entropy.  This conclusion is reached based upon three
considerations: that the dynamics of the system are Hamiltonian;
that a probabilistic description is meaningful; and that thermal
states are physically represented by canonical probability
distributions.

No assumptions were required regarding whether thermal systems are
subsystems of a large, microcanonically distributed system.
Consequently, no assumptions regarding ergodicity or mixing are
required.  No assumptions regarding the size of the systems are
involved, so no conclusions depend upon, or only hold true in, the
thermodynamic limit.  We have not assumed that the probability
distribution only applies to microscopic degrees of freedom. Should
probability distributions over macroscopically distinct
states\cite{Pen70} arise, the arguments still hold.  Of particular
importance, no restriction is made in its applicability to thermal
systems or systems in equilibrium. The arguments that identify the
Gibbs-von Neumann entropy for thermal systems, apply universally.

Beyond the use of a probabilistic description itself, no assumptions
were made regarding entropy having a relationship to knowledge or
information.  It is quite unnecessary to consider information theory
or properties of Shannon information.  No relationship between
thermal states and maximal ignorance need be assumed.  Describing
the Gibbs-von Neumann entropy as information theoretic seems
unjustified, if not downright anachronistic\footnote{The Gibbs
entropy appears in 1902\cite{Gib1902}and the generalisation to
quantum theory in 1932\cite{Neu55}.  The equivalent term does not
appear in information theory until 1948\cite{Sha48} and is not
generalised to quantum theory until 1994\cite{JS94}.}.

The physical understanding of the Gibbs-von Neumann entropy is shown
to be precisely the generalisation one should expect, to statistical
mechanics, of the thermodynamic entropy.  The generalisation is from
\begin{quote}
There is no process which takes a system from state A, to a system
at state B, depositing heats $Q_m$ into thermal systems at
temperatures $T_m$, for which
\begin{equation}
S\func{B}-S\func{A} + \sum_m \frac{Q_m}{T_m} < 0
\end{equation}
\end{quote}
to
\begin{quote}
There is no process which takes a system from state A, to a system
at state B, depositing \textit{ mean} heats $Q_m$ into thermal
systems at temperatures $T_m$, for which
\begin{equation}
S\func{B}-S\func{A} + \sum_m \frac{Q_m}{T_m} < 0
\end{equation}
\end{quote}
The generalisation involved recognises that, with some probability,
all of the classical statements of the second law of thermodynamics,
are violated to any degree.  The restriction expressed by the
entropy function is not of the minimal heat generation for a
Hamiltonian evolution of a given microstate, but of what is the
minimum \textit{ expectation value} of the heat generated by
Hamiltonian flows.

There remain many open questions in the understanding of statistical
mechanics\cite{Uff2006}.  We would like to develop how statistical
mechanics may account for them, but this paper is far too long
already.  Of particular importance is the exploration of how the
fine grained Gibbs-von Neumann entropy accounts for the appearance
of irreversibility and time-asymmetry.  On the question of whether
this entropy is subjective, and `observers' may affect the entropy
of a system, in the manner of a Maxwellian Demon, see\cite{Mar02}.
There are many other measures, microscopic and macroscopic, of
probability distributions and of individual states, that are
presented as `entropies'.  While they may have useful roles to play,
the question is: are they the statistical mechanical generalisation
of \textit{thermodynamic} entropy?

Once the canonical distribution is accepted as appropriate for
thermal states, the Gibbs-von Neumann entropy follows inevitably for
all probability distributions, microscopic or macroscopic.  In
Section \ref{s:temperature} it is shown that the canonical
distribution can be uniquely identified, solely from considering the
observed properties of thermal states themselves.  The relationship:
\begin{equation}
\sum_i \frac{\Delta Q_i}{T_i} \geq 0
\end{equation}
for closed cycles, is then a derived property of Hamiltonian
dynamics and serves to \textit{ define} the statistical mechanical
generalisation of entropy, in the same way that the Clausius
relationship defines phenomenological entropy. To quote a recent
paper\footnote{The views of which are, nevertheless, quite different
to the views of this paper!}:

\begin{quote}The rule  \dots  that associates heat transfer with entropy holds
only for thermodynamic entropy and, indeed, defines it.  No other
entropy can satisfy it without at once also being thermodynamic
entropy.\cite{Nor05}
\end{quote}

\textbf{Acknowledgements} The author wishes to thank Steve Weinstein
for his advice, John Norton and Jos Uffink for drawing my attention
to Szilard's 1925 paper, and Matt Leifer for discussions on passive
distributions.

Research at Perimeter Institute is supported in part by the
Government of Canada through NSERC and by the Province of Ontario
through MEDT.

\bibliographystyle{alpha}

\end{document}